# Symmetry-broken Chern insulators in twisted double bilayer graphene

Minhao He[1†], Jiaqi Cai[1†], Ya-Hui Zhang[2], Yang Liu[1], Yuhao Li[1], Takashi Taniguchi[3], Kenji Watanabe[4], David H. Cobden[1], Matthew Yankowitz[1,5#], and Xiaodong Xu[1,5#]

[1]Department of Physics, University of Washington, Seattle, Washington, 98195, USA
[2]Department of Physics and Astronomy, Johns Hopkins University, Baltimore, Maryland, 21218, USA
[3]International Center for Materials Nanoarchitectonics, National Institute for Materials Science, 1-1 Namiki, Tsukuba 305-0044, Japan
[4]Research Center for Functional Materials, National Institute for Materials Science, 1-1 Namiki, Tsukuba 305-0044, Japan
[5]Department of Materials Science and Engineering, University of Washington, Seattle, Washington, 98195, USA

[†] These authors contributed equally to the work.
[#]Correspondence to: myank@uw.edu; xuxd@uw.edu;

**Twisted double bilayer graphene (tDBG) has emerged as an especially rich platform for studying strongly correlated and topological states of matter. The material features moiré bands that can be continuously deformed by both perpendicular displacement field and twist angle. Here, we construct a phase diagram representing of the correlated and topological states as a function of these parameters, based on measurements on over a dozen tDBG devices encompassing the two distinct stacking configurations in which the constituent Bernal bilayer graphene sheets are rotated either slightly away from 0° or 60°. We find a hierarchy of symmetry-broken states that emerge sequentially as the twist angle approaches an apparent optimal value of $\theta \approx 1.34°$. Among them, we discover a sequence of symmetry-broken Chern insulator (SBCI) states that arise only within a narrow range of twist angles (≈1.33° to 1.39°). We observe an associated anomalous Hall effect at zero field in all samples supporting SBCI states, indicating spontaneous time-reversal symmetry breaking and possible moiré unit cell enlargement at zero magnetic field.**

The flat bands found in the growing family of graphene-based moiré materials offer a valuable platform for studying the interplay of strong correlations and topology[1,2]. Twisted double bilayer graphene (tDBG), composed of two slightly relatively rotated sheets of Bernal bilayer graphene, has been the focus of significant research attention[3–15] owing to its exceptional tunability with perpendicular displacement field[3–6], twist angle, and pressure[12]. Compared with twisted bilayer graphene (tBLG), in which strongly correlated states occur over a very narrow range near the magic angle of ≈1.1°, tDBG hosts correlated states over a wider range of twist angles due to its lack of a singular 'magic-angle' from theory. A wide variety of correlated phenomena have been reported already in tDBG, including states that are spin-polarized[4,5], valley-polarized[13], spin-valley polarized[15], and intervalley-coherent[16], as well as a correlated electron-hole state[8], the anomalous Hall effect (AHE)[14], and superconductivity in samples interfaced with $WSe_2$[17]. Despite this, the phase diagram with respect to twist angle has yet to be fully mapped, and it remains unclear whether there exist optimal twist angles for observing these phenomena.

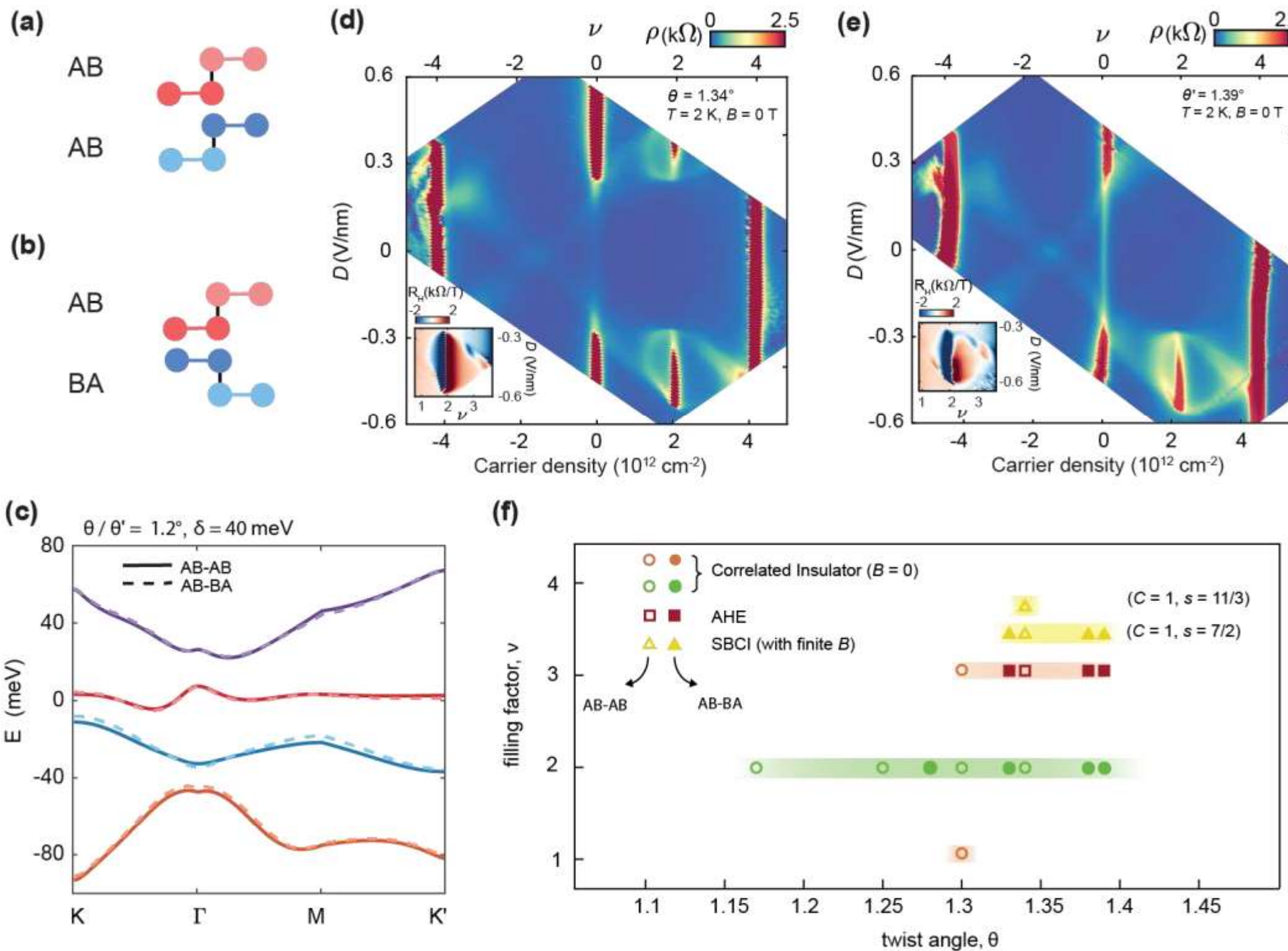

**Figure 1. Hierarchy of correlated and topological phases observed in tDBG in both stacking configurations. a-b,** Side-view lattice schematics of AB-AB and AB-BA tDBG. AB-AB (AB-BA) tDBG is composed of two Bernal stacked graphene bilayers with the same (opposite) orientation. **c**, Calculated band structure of $\theta$ = 1.2° (AB-AB, solid line) and $\theta'$ = 1.2° (AB-BA, dashed line) tDBG, with interlayer potential δ = 40 meV. The red bands denote the lowest moiré conduction band. The Chern number in the AB-AB (AB-BA) stacking is $C$ = 2 (1). **d-e,** Resistivity $\rho_{xx}$ maps of $\theta$ = 1.34° (AB-AB) and $\theta'$ = 1.39° (AB-BA) devices, respectively. Measurements were taken at $T$ = 2 K and zero magnetic field. Insets shows the zoomed-in Hall coefficient $R_{\mathrm{H}}$ at $|B|$ = 0.5 T, focusing on the correlated states. **f**, Summary of the correlated and topological states observed in 13 tDBG devices. Correlated insulating states, states exhibiting the AHE, and SBCI states are represented by circle, diamond, and triangle markers, respectively. The AB-AB (AB-BA) stacking is denoted by open (closed) markers. The green and orange shading denotes the range of twist angles over which symmetry-broken metallic states are observed, identified by sign reversals in the low-field Hall effect. States denoted by ($C$, $s$) correspond to SBCIs observed in a magnetic field.

In this work, we study over a dozen tDBG devices to systematically construct such a phase diagram. The devices span both stacking configurations of the two inner graphene sheets, distinguished by whether the two Bernal graphene sheets are rotated slightly away from 0° (AB-AB) or 60° (AB-BA) (see as indicated in Figs. 1a-b and Fig. S1). Previous studies have concentrated on the AB-AB case[17–20]. Here, we find that tDBG can host symmetry-broken Chern insulator (SBCI) states, likely connected to spontaneous translational symmetry breaking of the moiré unit cell. We additionally find that these states might be associated with an anomalous Hall effect at zero magnetic field. The observation of magnetic hysteresis at fractional band filling suggests that the translational symmetry breaking may persist down to zero field. Surprisingly, although our band structure calculations at zero field indicate that the Chern number should be different in the AB-AB and AB-BA cases (Fig. 1c, see also Fig. S2), we observe similar SBCI states at high field for both stacking configurations.

Our study included six AB-BA samples with $\theta$ varying from 1.06° to 1.39° and seven AB-AB samples with $\theta'$ varying from 1.17° to 1.53° (Table S1). Figures 1d and e show maps of the longitudinal resistivity, $\rho_{xx}$, versus carrier density, $n$, and displacement field, $D$ (see SI for definition) for AB-AB device S1 ($\theta$ = 1.34°) and AB-BA device O1 ($\theta'$ = 1.39°), respectively. The filling factor, $\nu$, is shown on the top axis, where $\nu = \pm 4$ corresponds to full filling of the lowest moiré conduction and valence minibands. The data is consistent with measurements on AB-AB tDBG detailed extensively elsewhere[18–22]. In brief, one sees gapped states at $\nu = 0$ and $\pm 4$ that are tuned by $D$, a cross-like resistive feature in the region $\nu < 0$ reflecting van Hove singularities in the valence band, and a 'halo' feature around the insulating states at $\nu = +2$ marking the emergence of symmetry-broken correlated metallic and insulating states. Measurements of the antisymmetrized Hall coefficient, $R_H = (R_{xy}[B] - R_{xy}[-B])/(2B)$, further reveal incipient symmetry-broken states at $\nu = +3$. All the same features are seen in the AB-BA devices (see also Fig. S3, S4), consistent with the nearly identical calculated band dispersions of the AB-AB and AB-BA types (Fig. 1c). Figure 1f summarizes the correlated and topological phases observed as a function of twist angle, which will be elaborated on below.

Our primary finding is the emergence of SBCI states in both AB-BA and AB-AB tDBG, as detailed in Figures 2 and 3, respectively. We first analyze the evidence for topological states in the AB-BA samples. Figures 2a and b show Landau fan diagrams of $\rho_{xx}$ and $\rho_{xy}$ acquired at $D = -0.40$ V/nm in device O1. We see signs of a variety of gapped states, each characterized by a sharp suppression of $\rho_{xx}$ and (nearly) quantized $\rho_{xy}$, that evolve along linear trajectories described by the equation $\nu = Cn_\phi + s$, where $n_\phi$ is the number of magnetic flux quanta per moiré cell, the integer $C$ is the Chern number, and $s$ is the band filling index (corresponding to the number of electrons per moiré unit cell)[23]. Figure 2c is a schematic showing the trajectories of the more robust gapped states. States with $C \neq 0$ and $s = 0$, equivalent to integer quantum Hall (IQH) states, are denoted in purple. States with $C \neq 0$ and integer $s$, known as Chern insulators (CI), are denoted in orange ($s$ = 1), red ($s$ = 2), and blue ($s$ = 3). Topologically trivial ($C$ = 0) insulating states are denoted in black.

The IQH and CI states seen here are all consistent with gapped states predicted by the Hofstadter model[23–27], and can be understood in the context of Hofstadter subband ferromagnetism[28–33]. However, we observe an additional robust gapped state with integer $C$ = 1 but with a fractional band filling index $s$ = 7/2 (denoted by the green dashed line in Fig. 2c). This state emerges abruptly above $B \approx 6$ T, and exhibits a quantization of the Hall resistivity at $\rho_{xy} = h/e^2$ (see also Fig. S5). Such a gapped state, arising at a partial filling of a Hofstadter band, can only be explained by incorporating the effects of strong Coulomb interactions going beyond simple isospin polarization. States with $C \in \mathbb{Z}$ but $s \notin \mathbb{Z}$ are most naturally associated with the formation of a SBCI[34–36], in which Coulomb interactions favor the spontaneous formation of a topological charge density wave state[20–22]. Given that this gapped state emerges upon doping beyond three electrons per moiré unit cell, the SBCI most likely forms within a single partially-occupied moiré subband with fully lifted isospin degeneracy. An $s$ = 7/2 SBCI state can arise owing to a spontaneous doubling of the area of the original moiré unit cell[37–39]. The corresponding density wave folds the moiré Brillouin zone, thereby doubling the number of moiré minibands. For a state with $s$ = 7/2, the SBCI corresponds to a complete filling of seven out of the eight available moiré conduction bands. Additional Landau fan diagrams (Figs. S6-7) show that the SBCI state persists over a relatively wide range of $D$. The same $C = 1$, $s = 7/2$ SBCI state was also observed in two other AB-BA samples with $\theta' = 1.33°$ and $\theta' = 1.38°$ (Fig. S8).

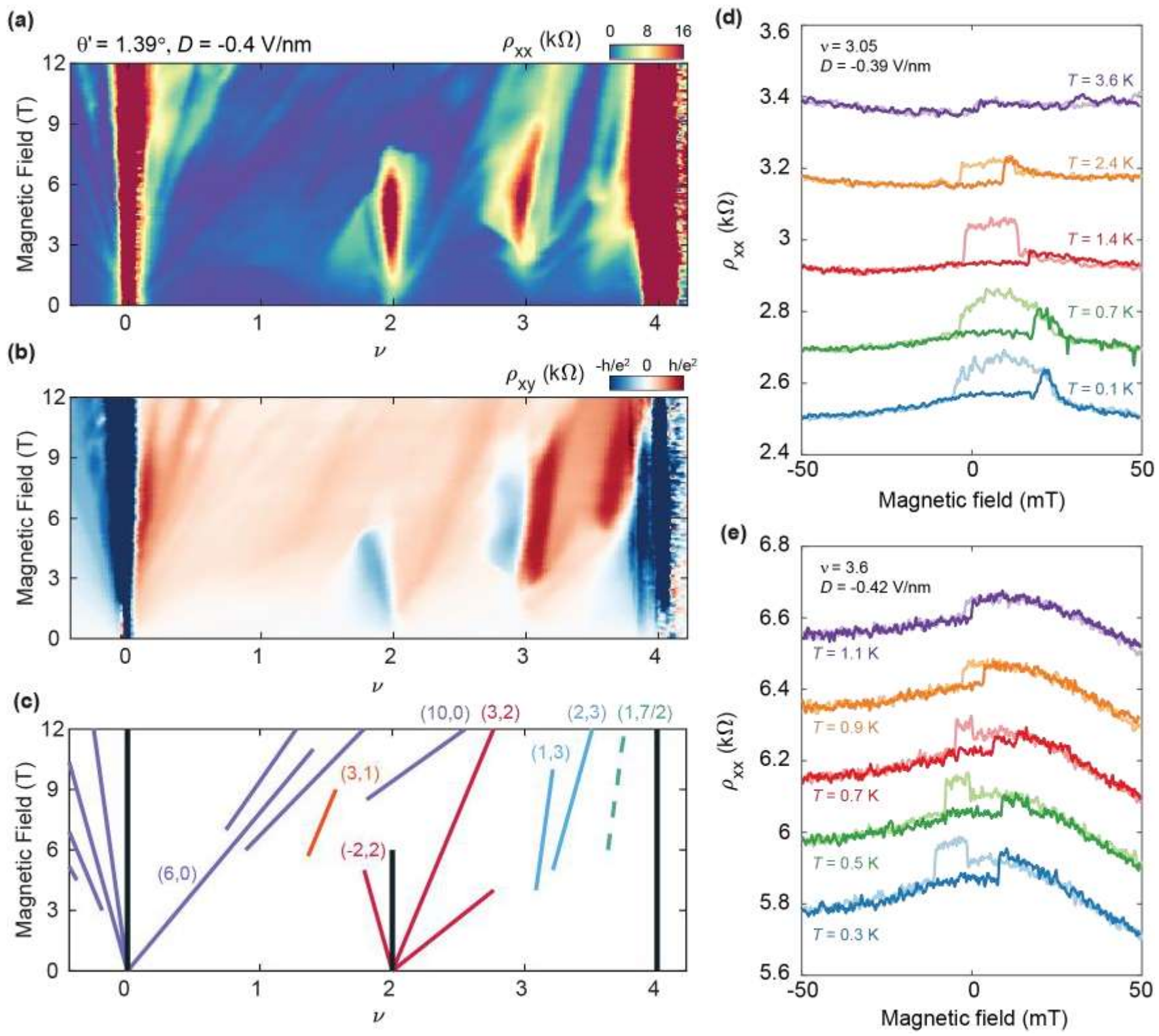

**Figure 2. Symmetry-broken Chern insulator and anomalous Hall effect. a-b,** Landau fan measurements of the longitudinal $\rho_{xx}$ and Hall resistivity $\rho_{xy}$ at $D$ = -0.4 V/nm in a $\theta'$ = 1.39° AB-BA sample. Data is taken at 100 mK. **c,** Schematics of all observed gapped states in **a-b**. Several main gapped states are labelled by their respective ($C$, $s$) values. Colors differentiate different values of s. Black vertical lines (with $C$ = 0) denote topologically trivial insulating states. The green dashed line denotes the (1, 7/2) symmetry-broken Chern insulator. **d-e,** $\rho_{xx}$ measured as magnetic field is swept back and forth at **d**: $\nu$ = 3.05, $D$ = -0.39 V/nm and **e**: $\nu$ = 3.6, $D$ = -0.42 V/nm. The data taken at different temperatures are offset vertically for clarity.

In addition to the SBCI at high magnetic field, we also see an AHE at zero field. Figure 2d shows measurements of $\rho_{xx}$ as the magnetic field is swept back and forth through zero at $\nu = 3.05$ and $D = -0.39$ V/nm at a series of temperatures $T$ (see also Fig. S9). The hysteresis seen at low $T$ at this filling implies orbital magnetism is present, and hence time-reversal symmetry is broken. Despite the nominally longitudinal contact geometry, our measurements show the square-loop–type behavior expected of a Hall geometry. This is likely the result of a mosaic of magnetic domains[40,41] in our sample resulting from structural disorder of the moiré pattern. Consistent with such a disordered domain picture, the details of the hysteresis change upon thermal cycling (Fig. S10a). The small amplitude of the AHE (much less than $h/e^2$) can also readily be explained by domains, although it is also possible that the ground state is only partially isospin polarized and not fully gapped. We also observed the AHE at $\nu = 3$ in two other AB-BA samples with $\theta' = 1.33°$ and 1.38° (Fig. S10). Notably, the AHE can sometimes be seen even far from $\nu = 3$, as illustrated in Fig. 2e for $\nu = 3.61$ (see also Fig. S11). This AHE near $\nu = 7/2$ suggests that the topological charge density wave state responsible for the SBCI at high field may still be present at zero field (see detailed discussion in the Supporting Information and Figs. S12-14). Similar observations have been reported in twisted monolayer-bilayer graphene[36].

Returning to the AB-AB case (Fig. 3), we find a similar SBCI with $C = 1$ and $s = 7/2$ emerging abruptly above $B = 5$ T, as seen in the Landau fan diagram taken at $D = -0.34$ V/nm in device S2 (Figs. 3a-b). Figure 3c is a schematic of the trajectories of the gapped states in Fig. 3a, which can be compared with Fig. 2c. Here, however, a second $C = 1$ SBCI emerges abruptly above 9 T with fractional band filling index $s = 11/3$ (see also Fig. S15). Analogous to the reasoning for the 7/2 state above, such a state can arise upon filling 11 of the 12 moiré subbands in an enlarged unit cell with triple the moiré unit cell area. Figure 3d shows a line cut of $\rho_{xx}$ and $\rho_{xy}$ at B = 10 T, where the 7/2 and 11/3 SBCI states (indicated by green shading) are both stable. Due to the small thermal activation gap of the $s = 11/3$ SBCI (Figure 3d inset and Fig. S16-17), the Hall resistivity $\rho_{xy}$ only reaches about 80% of $h/e^2$ at 100 mK. The $s = 7/2$ and 11/3 states should be distinguishable based on their different electron density modulations in real space which could be observable in scanning tunneling microscopy experiments.

Notably, the $s = 7/2$ and 11/3 states both have a Chern number of $C = 1$ despite the presumed difference in their unit cell areas. Both appear to emerge out of a parent state with $C = 2$, determined by identifying the most robust of the $s = 3$ states in the Landau fan diagram (Figs. 3a-c). A $C = 2$ Chern band can be mapped onto a two-component quantum Hall system. In such a model, the $s = 7/2$ SBCI can be understood as a spontaneously polarized state of this two-component system; essentially, a lattice analogue of the typical quantum Hall ferromagnet[37–39]. In a band with $C = 2$, the Berry curvature can partition evenly into each of the two components, consistent with the observed $C = 1$ of the SBCI when the degeneracy is lifted. However, the $s = 11/3$ state is more complicated in this model, since a two-component $C$ = 2 state cannot be evenly partitioned into three subbands with integer Chern numbers. Such a state thus requires either a non-uniform partitioning of the Chern band into different subbands, or a dynamical renormalization of the Chern number of the parent subband upon doping (i.e., from $C$ = 2 to $C$ = 3). Therefore, the question of how the $s$ = 11/3 state has $C$ = 1 requires further theoretical attention.

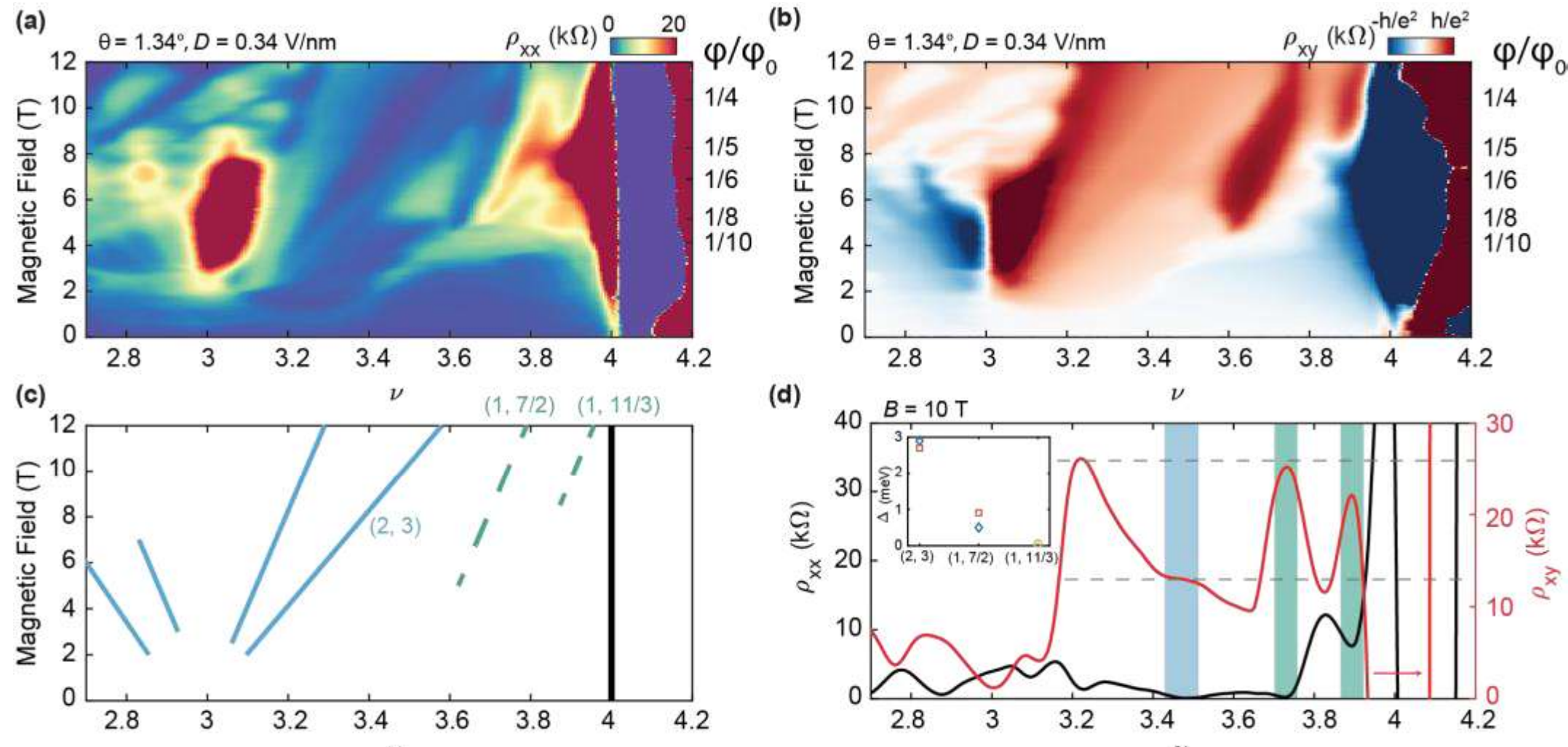

**Figure 3. Hierarchy of SBCIs with different translational symmetry breaking. a-b,** Landau fan diagram of the longitudinal $\rho_{xx}$ and Hall resistivity $\rho_{xy}$ at $D$ = 0.34 V/nm in a $\theta$ = 1.34° AB-AB tDBG sample. Data is taken at 100 mK. **c,** Schematics of all observed gapped states in **a-b**. Several main gapped states are labelled by their respective $(C, s)$ values, using the same color coding as in Fig. 2. The green dashed lines denote the (1, 7/2) and the (1, 11/3) SBCIs . **d,** Linecuts of $\rho_{xx}$ and $\rho_{xy}$ from panels **a-b** at $B$ = 10 T. The grey dashed line corresponds to $h/e^2$ and $h/2e^2$ for $\rho_{xy}$. The SBCIs are emphasized by the green shadings. The nearby quantized (2, 3) Chern insulator is marked by the blue shading. The inset shows the thermal activation gap of the (2,3), (1, 7/2) and (1, 11/3) states. The blue diamonds denote data from the $\theta'$ = 1.39° AB-BA sample at $D$ = -0.4 V/nm with $B$ = 8 T. The orange square denote data from the $\theta$ = 1.34° AB-AB sample at $D$ = 0.4 V/nm with $B$ = 9 T. The yellow circle denotes data from the same $\theta$ = 1.34° AB-AB sample at $D$ = 0.34 V/nm with $B$ = 9 T.

We conclude by returning to the summary of the correlated and topological states observed in our 13 tDBG devices shown in Fig. 1f. Overall, we see a clear pattern of states emerging within well-defined ranges of twist angle, symmetric with respect to $D$ field orientations except for the one device additionally aligned with h-BN (see discussion on the aligned case in SI, and Fig. S18-19). There is little apparent dependence on the stacking type, AB-AB or AB-BA, though we note that it is in principle possible to misidentify the stacking order of a given sample in the case where a strain soliton separated the two stacked regions of bilayer graphene in the exfoliated flake[42], unintentionally reversing the intended stacking order. The symmetry-broken state at $\nu$ = 2 is the most robust, seen in devices with twist angles between 1.17° and 1.41°. An additional symmetry-broken state occurs in the vicinity of $\nu$ = 3 for twist angles between 1.30° and 1.39°, with an incipient insulating state seen in a single device[7] with $\theta = 1.30°$ (see also Ref. 14). Similarly, a symmetry-broken state at $v = 1$ only emerges at zero field in the device with $\theta = 1.30°$. All devices with twist angles between 1.33° and 1.39° exhibit the AHE at $B = 0$ and at least one SBCI at high magnetic field. The device with $\theta = 1.34°$ also exhibits an $s = 11/3$ SBCI. Taken together, our observations suggest that the richest set of correlated and topological phases are found in tDBG samples with twist angles within the range ≈1.30°-1.39°. Therefore, although flat bands in tDBG can be formed over an extended range of twist angles owing to the tunability of the bands with displacement field, we have identified a narrow range of twist angles over which the strength of correlations appears to be greatest.

**Acknowledgments:** This research was primarily supported by the U.S. National Science Foundation through the UW Molecular Engineering Materials Center (MEM-C), a Materials Research Science and Engineering Center (DMR-1719797 and DMR-2308979), including the use of facilities and instrumentation provide by the MEM-C Shared Facilities. M.H. and M.Y. acknowledge support from the Army Research Office under Grant Number W911NF-20-1-0211. Y.H.Z. was supported by the National Science Foundation under Grant No. DMR-2237031. X.X. and M.Y. acknowledge support from the State of Washington funded Clean Energy Institute. This work made use of a dilution refrigerator system which was provided by NSF DMR1725221. K.W. and T.T. acknowledge support from the Elemental Strategy Initiative conducted by the MEXT, Japan, Grant Number JPMXP0112101001, JSPS KAKENHI Grant Number JP20H00354 and the CREST (JPMJCR15F3), JST.

**Author Contributions:** X.X. and M.Y. conceived and supervised the experiment. M.H. and J. C. fabricated the devices and performed the measurements. Y.Z. performed the calculations. Y. Liu. and Y. Li. contributed to fabrication of AB-AB stacked devices. D.H.C. assisted with measurements in the dilution refrigerator. K.W. and T.T. provided the bulk hBN crystals. M.H., J.C., M.Y., and X.X. analyzed the data and wrote the paper with input from all authors.

**Competing Interests:** The authors declare no competing interests.

**Data Availability**: All data that support the plots within this paper and other findings of this study are available from the corresponding author upon reasonable request. Source data are provided with this paper.

# Supporting Information for: Symmetry-broken Chern insulators in twisted double bilayer graphene

Minhao He,[†,⊥] Jiaqi Cai,[†,⊥] Ya-Hui Zhang,[‡] Yang Liu,[†] Yuhao Li,[†] Takashi Taniguchi,[¶] Kenji Watanabe,[§] David H. Cobden,[†] Matthew Yankowitz,[*,†,‖] and Xiaodong Xu[*,†,‖]

†*Department of Physics, University of Washington, Seattle, Washington, 98195, USA*

‡*Department of Physics and Astronomy, Johns Hopkins University, Baltimore, Maryland, 21218, USA*

¶*International Center for Materials Nanoarchitectonics, National Institute for Materials Science, 1-1 Namiki, Tsukuba 305-0044, Japan*

§*Research Center for Functional Materials, National Institute for Materials Science, 1-1 Namiki, Tsukuba 305-0044, Japan*

‖*Department of Materials Science and Engineering, University of Washington, Seattle, Washington, 98195, USA*

⊥*These authors contributed equally to this work.*

E-mail: myank@uw.edu; xuxd@uw.edu

# Methods

**Device fabrication.** tDBG devices were fabricated using a cut-and-stack method. Bilayer graphene is first cut using an AFM lithography technique, and the two pieces are picked up sequentially while setting the twist angle to the desired value close to 60° in between. We use 3-5 nm graphite as both top and bottom gates. Heterostructures of graphite/hBN/tDBG/hBN/graphite are assembled using a standard dry-transfer technique with a PC/PDMS (polycarbonate/polydimethylsiloxane) stamp and transferred onto a Si/$SiO_2$ wafer. $CHF_3/O_2$ etching and $O_2$ plasma etching followed by electron beam lithography are used to define a Hall bar geometry, and Cr/Au contacts are finally added using electron beam evaporation. A subset of the AB-AB tDBG devices used in the current study are the same as those reported in our previous work.[1]

We note that our assignment of sample stacking configuration is based on the assumption of single domain Bernal bilayer graphene used in fabrication. We label the devices by their nominal stacking configuration. We fabricate the Hall bar devices with typical contact pair size about 1 $\mu$m by 2 $\mu$m, to reduce possible artifacts from domain walls.

**Transport measurements.** Transport measurements at temperatures above 2 K were conducted in a Quantum Design PPMS system. Transport measurements below 2 K were conducted in a Bluefors dilution refrigerator with cryogenic low-pass filters. In both cases, measurements are performed with a 1-5 nA a.c. excitation current at either 13.3 Hz or 13.7 Hz. The current and voltage are pre-amplified by DL 1211 and SR560 respectively, and then read out by SR830/SR860 lock-in amplifiers. Gate voltages are supplied by either NI DAQ or Keithley 2450. A global Si gate is sometimes used to reduce contact resistance issue. The dual-gate device geometry enables independent control of D and n. The relationship between the top and bottom gate voltages, $V_t$, $V_b$, and $n$ and $D$ are given by $n = (V_tC_t + V_bC_b)/e$ and $D = (V_tC_t - V_bC_b)/2\epsilon_0$, where $C_t$ and $C_b$ are the top and bottom gate capacitances per unit area and $\epsilon_0$ is the vacuum permittivity. We note that our definition of $D$ follows the convention in which it has the same units as electric field.

The filling factor $\nu$ is defined as the number of electrons per moiré unit cell. Full filling of the four-fold spin-valley degenerate flat bands corresponds to 4 electrons (holes) per moiré unit cell, $\nu = +4$ (-4). The twist angle was first determined by measuring the carrier density corresponding to the band insulators at $\nu = \pm 4$, following the relationship $n = 8\theta^2/\sqrt{3}a^2$, where a = 0.246 nm is the graphene lattice constant. It is further confirmed by fitting the observed quantum Hall states and Chern insulators with the allowed Hofstadter states in the Wannier diagram. Brown-Zak oscillations were also used to verify the twist angle whenever applicable. Our devices have a typical twist angle disorder of around 0.02° between contact pairs.

**Band structure calculation of AB-BA tDBG.** We calculate the band structure of the AB-BA stacked tDBG based on the standard Bistritzer-MacDonald continuum model. The full Hamiltonian for the valley $K$ is:

$$H = H_t^{AB} + H_b^{BA} + H_M \tag{S1}$$

where $H_t^{AB}$ and $H_b^{BA}$ are the Hamiltonian of the two layers separately. $H_M$ includes the interlayer moiré tunneling terms. We have

$$H_t^{AB} = \Psi_t^\dagger \begin{pmatrix} \frac{D}{2} & t(k_x - ik_y)e^{i\frac{\theta}{2}} & t_4(k_x - ik_y)e^{i\frac{\theta}{2}} & t_3(k_x + ik_y)e^{-i\frac{\theta}{2}} \\ t(k_x + ik_y)e^{-i\frac{\theta}{2}} & \frac{D}{2} & \gamma_1 & t_4(k_x - ik_y)e^{i\frac{\theta}{2}} \\ t_4(k_x + ik_y)e^{-i\frac{\theta}{2}} & \gamma_1 & \frac{D}{6} & t(k_x - ik_y)e^{i\frac{\theta}{2}} \\ t_3(k_x - ik_y)e^{i\frac{\theta}{2}} & t_4(k_x + ik_y)e^{-i\frac{\theta}{2}} & t(k_x + ik_y)e^{-i\frac{\theta}{2}} & \frac{D}{6} \end{pmatrix} \Psi_t \tag{S2}$$

and

$$H_b^{BA} = \Psi_b^\dagger \begin{pmatrix} -\frac{D}{6} & t(k_x - ik_y)e^{-i\frac{\theta}{2}} & t_4(k_x + ik_y)e^{i\frac{\theta}{2}} & \gamma_1 \\ t(k_x + ik_y)e^{i\frac{\theta}{2}} & -\frac{D}{6} & t_3(k_x - ik_y)e^{-i\frac{\theta}{2}} & t_4(k_x + ik_y)e^{i\frac{\theta}{2}} \\ t_4(k_x - ik_y)e^{-i\frac{\theta}{2}} & t_3(k_x + ik_y)e^{i\frac{\theta}{2}} & -\frac{D}{2} & t(k_x - ik_y)e^{-i\frac{\theta}{2}} \\ \gamma_1 & t_4(k_x - ik_y)e^{-i\frac{\theta}{2}} & t(k_x + ik_y)e^{i\frac{\theta}{2}} & -\frac{D}{2} \end{pmatrix} \Psi_b \tag{S3}$$

In the above, we choose basis as $\Psi_{t/b} = (f_{A_{t/b;1}}, f_{A_{t/b;2}}, f_{B_{t/b;1}}, f_{B_{t/b;2}})^T$ and $f_{A_{t/b;1}}, f_{A_{t/b;2}}, f_{B_{t/b;1}}, f_{B_{t/b;2}}$ label the layer-sublattice degree of freedom of the top/bottom bilayer graphene. We use the parameters $t = -3100\frac{\sqrt{3}}{2}$ meV, $t_1 = 283\frac{\sqrt{3}}{2}$ meV, $t_4 = 110\frac{\sqrt{3}}{2}$ meV and $\gamma_1 = 361$ meV, where the momentum $k_x, k_y$ is in the unit of $\frac{1}{a}$, and $a \approx 0.246$ nm is the lattice constant of graphene.

The interlayer moiré tunneling Hamiltonian $H_M$ is

$$H_M = \sum_{j=1,2,3} (f^\dagger_{A_{t;2}}(\mathbf{k}), f^\dagger_{B_{t;2}}(\mathbf{k})) T_j \begin{pmatrix} f_{A_{b;1}}(\mathbf{k} + \mathbf{Q_j}) \\ f_{B_{b;1}}(\mathbf{k} + \mathbf{Q_j}) \end{pmatrix} + h.c. \tag{S4}$$

where $\mathbf{Q_1} = 0$, $\mathbf{Q_2} = \mathbf{G_1} = (-\frac{2\pi}{\sqrt{3}a_M}, -\frac{2\pi}{a_M})$, and $\mathbf{Q_3} = \mathbf{G_2} = (\frac{2\pi}{\sqrt{3}a_M}, -\frac{2\pi}{a_M})$. The moiré lattice constant at twist angle $\theta$ is denoted as $a_M = \frac{a}{2\sin\frac{\theta}{2}}$. We have $T_j = t_M \begin{pmatrix} \alpha & e^{-i\frac{2\pi j}{3}} \\ e^{i\frac{2\pi j}{2}} & \alpha \end{pmatrix}$, where $t_M$ is the interlayer tunneling strength and $\alpha \in [0, 1]$ is introduced to incorporate the effects of lattice relaxation. We use $t_M = 110$ meV and $\alpha = 0.8$.

At the limit of $D = 0$, there is a mirror symmetry $M : k_x \to k_x, k_y \to -k_y$, which acts as $M_y K$, where $K$ is the complex conjugate. $M_y$ maps $A_{t;1} \to A_{b;2}, A_{t;2} \to A_{b;1}, B_{t;1} \to B_{t;2}, B_{t;2} \to B_{b;2}$, where the momentum changes as $k_x \to -k_x, k_y \to k_y$. There is no flip of the valley index in this transformation. We note that the realization of the mirror symmetry is different for the AB-AB stacked tDBG, in which $M : k_x \to k_x, k_y \to -k_y$, acting as $M_x$, with $A_{t;1} \to B_{b;2}, A_{t;2} \to B_{b;2}, B_{t;1} \to A_{b;2}, B_{t;2} \to A_{t;2}$. Again there is no valley flip in the transformation. Note that there is no complex conjugate compared to the AB-BA stacking.

As a result, the constraint of the Chern number $C$ is different. For AB-AB stacking, we have $C(D) = -C(-D)$ within one valley, while we have $C(D) = C(-D)$ within each valley for the AB-BA stacking.

The band dispersions of AB-AB tDBG and AB-BA tDBG are quite similar.[2–6] We calculate the band structure of tDBG in both stacking chiralities at $\theta(\theta') = 1.2^{\circ}$ with a finite $D$, represented by interlayer potential $\delta = 40$ meV. Although the band structures are almost identical (Fig. 1c of the main text), the Berry curvature distributions are very different, as shown in Fig. S2. For AB-AB stacked tDBG, the Berry curvature from the K and K' points of the moiré Brillouin zone are the same and sum together to result in $C = 2$ for the lowest conduction band. In contrast, for AB-BA tDBG, $K$ and $K'$ carry opposite Berry curvatures. The net Berry curvature comes primarily from the center region of the moiré Brillouin zone (*i.e.*, the $\Gamma$ point), resulting in $C = 1$.

## S1. AHE near fractional filling $\nu = 7/2$

We discuss the nature of the AHE shown in Figure 2e of the main text. We first examine the doping dependence of the AHE to distinguish its relation with the AHE observed exactly at $\nu = 3$. We map the difference in $\rho_{xy}$ between the two field sweep directions, $\Delta\rho_{xy} = \rho_{xy}^{B\downarrow} - \rho_{xy}^{B\uparrow}$ with $D = -0.42$ V/nm (Fig. S12). We observe two distinct pockets of clear AHE, one around $\nu = +3$ and one around $\nu = +3.5$. We also find that the optimal $D$ for the AHE is different for the states around $\nu = +3$ and $\nu = +3.5$ (Fig. S13b). While the two pockets appear to merge and extend over a wide range of $\nu$ for some specific values of $D$ (Fig. S13a), their typical separation in both doping and displacement field suggests that they might be associated with distinct correlated ground states. Correlated states at non-integer filling of a moiré miniband have been observed previously in graphene aligned with hBN at large magnetic field,[7,8] and more recently in twisted monolayer-bilayer graphene (tMBG)[9] and twisted bilayer graphene (tBLG) aligned with hBN at $B = 0$.[10] Similarly, the observed AHE at fractional filling in tDBG may be associated with a spontaneous charge density wave (CDW) formation that expands the effective unit cell.

To explore the possibility of such spontaneous CDW formation, we performed a modified continuum model calculation generalized to include a CDW order (see detail in Note S5 below). This model predicts that the lowest moiré conduction band splits into two gapped subbands upon inclusion of the density wave order (Fig. S14), with a corresponding electron density profile shown in Fig. S14c. After considering remote band mixing, we calculate a valley Chern number of $C_v = 1$ for the higher-energy subband, consistent with our observation of a topological correlated state near $\nu = 7/2$.

## S2. Topological charge density wave states

We calculate reconstructed band structure and Chern number at $\nu = \frac{1}{2}$ after adding a phenomenological CDW order with $2 \times 1$ Moiré unit cell in the AB-BA stacked tDBG.

The CDW is generated by a potential with nesting momentum $\mathbf{Q} = (\frac{2\pi}{\sqrt{3}a_M}, 0)$. The interaction Hamiltonian at this momentum is given as:

$$H_V = V(\mathbf{Q})\rho(\mathbf{Q})\rho(-\mathbf{Q}) \tag{S5}$$

where $\rho(\mathbf{Q})$ is the density operator projected to the active band and $V(\mathbf{Q})$ is the Fourier-transformed screened Coulomb potential.

The mean field Hamiltonian decoupled from the interaction is:

$$H = H_{\text{nonint}} + (\Phi\rho(\mathbf{Q}) + h.c.) \tag{S6}$$

where, $H_{\text{nonint}}$ is non-interacting Hamiltonian defined in Eq.1, $\Phi = V(\mathbf{Q})\langle\rho(\mathbf{Q})\rangle$ is the charge density wave order parameter. In a more detailed form:

$$H = H_{\text{nonint}} + \sum_{\mathbf{k}} (\Phi\Lambda_{ab}(\mathbf{k}, \mathbf{Q}) + \Phi^*\Lambda_{ab}(\mathbf{k}, -\mathbf{Q}))c_a^\dagger(\mathbf{k}+\mathbf{Q})c_b(\mathbf{k}) + h.c. \tag{S7}$$

where $\Lambda_{ab}(\mathbf{k}, \mathbf{q}) = \langle\mu_a(\mathbf{k}+\mathbf{q})|\mu_b(\mathbf{k})\rangle$ is a form-factor entering the density operator. $\mu_a(\mathbf{k})$ is the Bloch wavefunction for band $a$.

Note here $\Phi$ can be a complex number, whose phase $\phi$ determines the phase of the CDW order $\rho(\mathbf{r}) = \rho_0 + A\cos(\pi x + \phi)$, where $x$ is the coordinate along the $\mathbf{a}_1$ direction. We label the AA site (maximum of the LDOS) as $\mathbf{r} = 0$ and choose $\phi = 0$ to pin the CDW with the moiré lattice (Fig. S14c).

In the original electron operator,

$$H = H_{\text{nonint}} + \sum_{\mathbf{k}} \sum_{a,b} (\Phi \Lambda_{ab}(\mathbf{k}, \mathbf{Q}) + \Phi^* \Lambda_{ab}(\mathbf{k}, -\mathbf{Q})) \mu_{a;\alpha}(\mathbf{k}+\mathbf{Q}) \mu^*_{b;\beta}(\mathbf{k}) f^\dagger_\alpha(\mathbf{k}+\mathbf{Q}) f_\beta(\mathbf{k}) + h.c. \tag{S8}$$

where $\alpha, \beta$ is a combination of the sublattice-layer index and the $m\mathbf{G}_1 + n\mathbf{G}_2$ index. $a, b$ is the band index. We sum $a, b$ with $M$ to be the band index cutoff, *i.e.*, over the conduction band, $M$ bands below and $M$ bands above. If we set $M = 0$, then the CDW only influences the conduction band. If $M \geq 1$, the CDW order also acts on the remote bands and couples the remote band to the conduction band. In practice, the result converges quite rapidly when increasing $M$. However, we note that $M \geq 1$ and $M = 0$ can give different results. The CDW order also hybridizes the remote band and the conduction band, which can renormalize the Chern number of the upper band split from the remainder of the conduction band.

We consider tDBG in both stacking chiralities at the same condition discussed above. In AB-AB tDBG, it is natural to expect that a CDW with $2 \times 1$ unit cell will split the original $C = 2$ conduction band into two $C = 1$ and $C = 1$ subbands. However, for AB-BA stacking, the $C = 1$ conduction band will split into a $C = 0$ and $C = 1$ band: the half around the zone center region carries $C = 1$, whereas the half around the $K$ and $K'$ carries $C = 0$. We calculate band structure in AB-BA tDBG with a CDW order $\Phi = 16$ meV in Fig. S14. We find that the Chern numbers of the gapped subbands are sensitive to the remote band renormalization. For $M = 0$, the upper subband has $C = 0$, inconsistent with our experimental observations of a topological state at $\nu = \frac{7}{2}$. After considering $M = 4$ remote bands, the Chern number of the upper subband renormalizes to $C = 1$, in agreement with our experiment. We note that $M = \infty$ (i.e. the realistic limit) should yield the same result for the conduction band as $M = 4$ in our calculation owing to its rapid convergence.

## S3. Relationship between the AHE and the high-field correlated Chern insulators

Near $\nu = 3$ and 3.5, we observe the AHE around $B = 0$ but a slow transition to a quantized Chern insulator state that only emerges at high field. Fig. S5 shows the development of the gapped states with field. Taken at face value, these measurements imply that the AHE surrounding zero field near $\nu = 3$ and 3.5 are not directly connected with the high-field Chern insulator and SBCI states. We propose two possible scenarios: (i) there is a magnetic field-induced phase transition between the zero-field and high-field symmetry broken states; (ii) the two arise from the same correlated state, but the quantization at low fields is obscured by moiré disorder and/or competition with quantum oscillations corresponding to symmetry-broken states emerging from nearby filling factors (e.g., the $\nu = 3.5$ state may compete with the quantum oscillations emerging from the $\nu = 3$ state). Compressibility measurements such as capacitance and scanning SET could help to resolve this question, particularly those offering local imaging which could probe length scales smaller than the moiré disorder.

## S4. Robust SBCI with hBN alignment

In Figure S18, we investigate the robustness of the SBCI state by studying a AB-BA sample ($\theta = 1.38°$) that is rotationally aligned with the top boron nitride (hBN) (device O2). Measurements of the Landau fan diagram at $D = 0$ V/nm confirm the existence of a second moiré potential arising from a twist angle of 0.55° between the uppermost graphene sheet and the encapsulating hBN (Fig. S19). This aligned hBN layer breaks the equivalence of the band structure for opposite orientations of $D$. As a result, the correlated states depend on the sign of $D$ (Fig. S18a), unlike in the hBN-misaligned samples. For positive $D$, electrons are pushed away from the graphene layer at the aligned hBN interface (Fig. S18b) while for negative D they are pushed towards it (Fig. S18c). For positive $D$, the correlated insulating state at $\nu = 2$ resembles that in hBN-misaligned samples, while for negative $D$ the correlated states are more complicated and exhibit new splitting features. The right-most panel of Fig. S18a shows the temperature dependence of $\rho_{xx}$ as a function of $D$ at $\nu = 2$. For $D < 0$ the correlated insulating state at $\nu = 2$ appears weakened, witnessed by the suppression of $\rho_{xx}$. Despite this, we still observe a robust C = 1, s = 7/2 SBCI state upon applying a magnetic field, albeit with a larger onset field than for $D > 0$ (green dashed lines in Fig. S18d-e). These observations emphasis the robustness of the SBCI state, which survives even upon weakening the correlation strength.

Table S1: Correlated and topological states observed in 13 tDBG samples are summarized. Samples are presented in the order of ascending twist angle and labelled with their nominal stacking configuration. Absence of the correlated and topological states are labelled specifically by None while / labels lack of proper measurements at the mK temperature range. Here, CI denote correlated insulator, halo denote symmetry breaking metallic state in the halo shape, AHE denote anomalous Hall effect with spontaneous time-reversal symmetry breaking, specifically at filling factor $\nu = 3$. Note that AHE near fractional filling $\nu = 3.5$ is observed in the $\theta' = 1.39°$ AB-BA stacked sample solely. $(C, s)$ label the Chern number and band filling index of the observed symmetry-broken Chern insulators.

| Twist angle | Stacking configuration | $\nu = 1$ | $\nu = 2$ | $\nu = 3$ | AHE at $\nu = 3$ | SBCI |
|---|---|---|---|---|---|---|
| 1.06 | AB-BA | None | None | None | / | / |
| 1.17 | AB-BA | None | halo only | None | / | None |
| 1.17 | AB-AB | None | CI with halo | None | / | / |
| 1.25 | AB-AB | None | CI with halo | None | / | / |
| 1.28 | AB-BA | None | CI with halo | None | / | None |
| 1.30 | AB-AB | CI with halo | CI with halo | CI with halo | No AHE | None |
| 1.33 | AB-BA | None | CI with halo | Resistive state with halo | AHE | (1, 7/2) |
| 1.34 | AB-AB | None | CI with halo | Halo only | / | / |
| 1.34 | AB-AB | None | CI with halo | Halo only | AHE | (1, 7/2) , (1, 11/3) |
| 1.38 (h-BN aligned) | AB-BA | None | CI with halo | Halo only | AHE | (1, 7/2) |
| 1.39 | AB-BA | None | CI with halo | Halo only | AHE | (1, 7/2) |
| 1.41 | AB-AB | None | Resistive state with halo | None | No AHE | None |
| 1.53 | AB-AB | None | None | None | / | / |

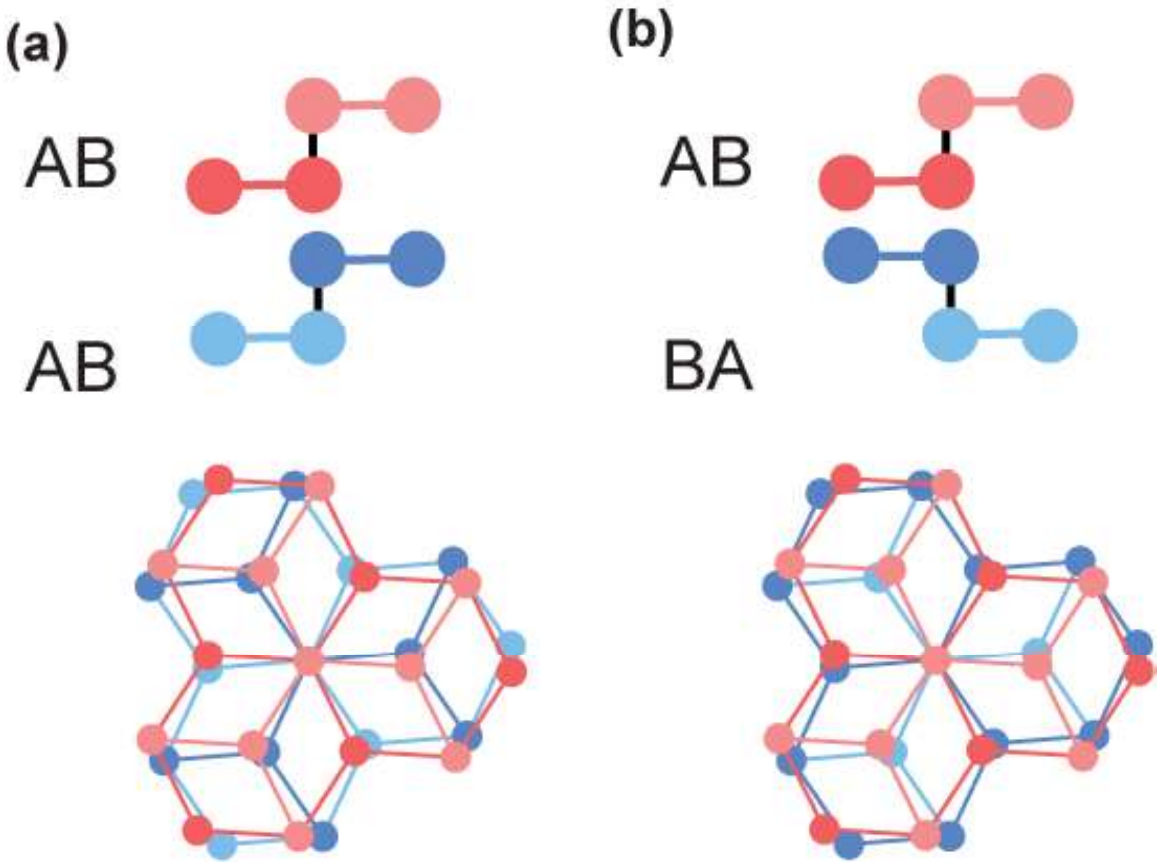

Figure S1: **Schematic drawings of AB-AB and AB-BA tDBG. a-b**, Lattice schematics of AB-AB and AB-BA tDBG, respectively. The component Bernal graphene bilayers are rotated near 0° and 60° in the two cases. The top row shows the side view while the bottom row shows the top view, focusing on the high symmetry local registry.

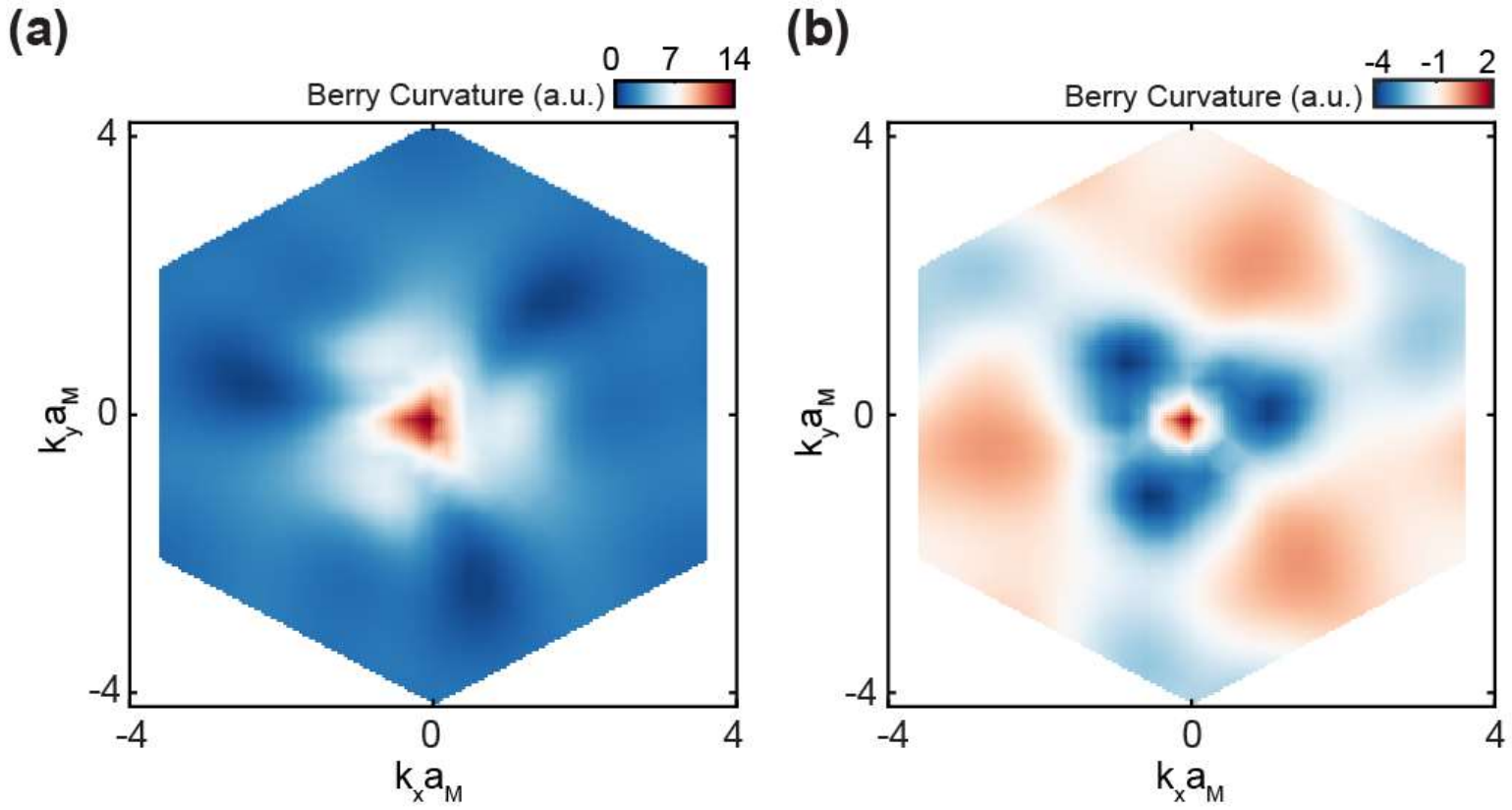

Figure S2: **Berry curvature of the lowest moiré conduction band for tDBG.** Corresponding Berry curvature ($\Omega$) of the lowest moiré conduction band for AB-AB and AB-BA, repectively, with $\delta = 40$ meV. $k$ is in the unit of $1/a_M$, where $a_M$ is the moiré lattice constant.

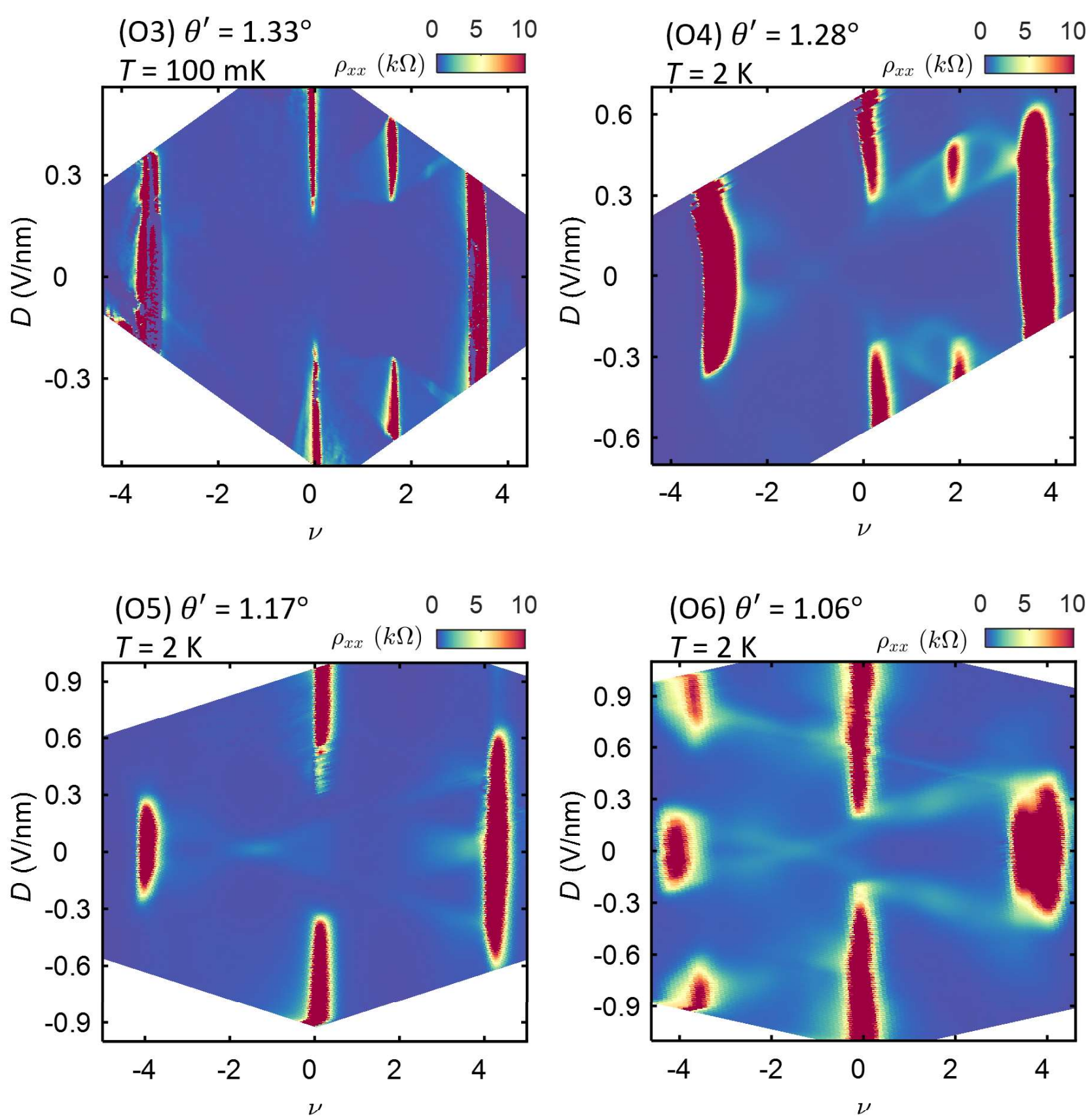

Figure S3: **Low temperature transport in additional AB-BA tDBG devices.** The twist angle and measurement temperature are denoted at the top left corner of each map. We observe displacement field-tunable insulating states at $\nu = 0$ and $\pm 4$ in all devices. Correlated insulating states appear clearly in devices O3 and O4.

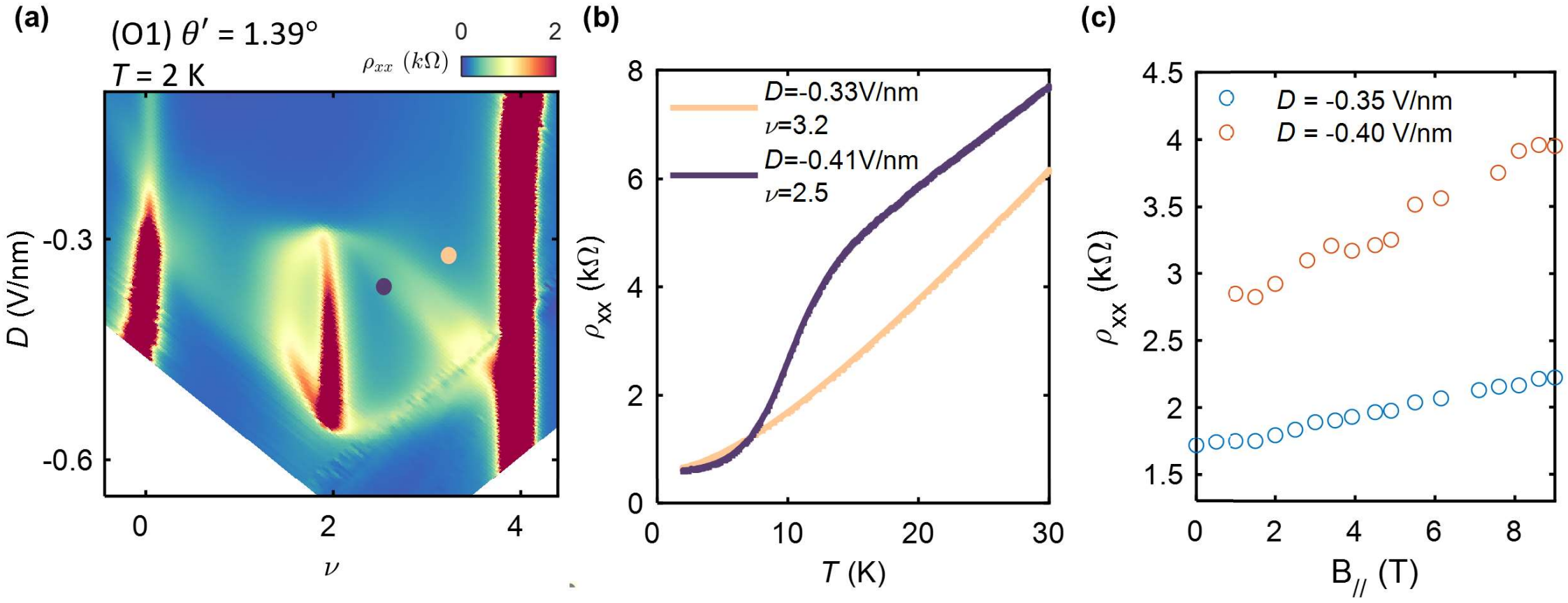

Figure S4: **Correlated phases in a $\theta' = 1.39^\circ$ AB-BA tDBG device (O1). a**, $\rho_{xx}$ map of device O1 at $T = 2$ K. **b**, $\rho_{xx}$ as a function of temperature inside (purple) and outside (yellow) the symmetry-broken metallic regime. $\rho_{xx}$ drops abruptly with temperature inside the symmetry-broken regime (purple curve), similar to observations in AB-AB tDBG.[1,11–14] This behavior is not observed outside the 'halo'-like region (orange curve). **c**, $\rho_{xx}$ as a function of in-plane magnetic field, $B_{||}$, at the $\nu = 2$ correlated insulating state with different $D$. We observe a weak increase of $\rho_{xx}$, suggesting that it may be spin-polarized as in AB-AB tDBG.[1,11–14]

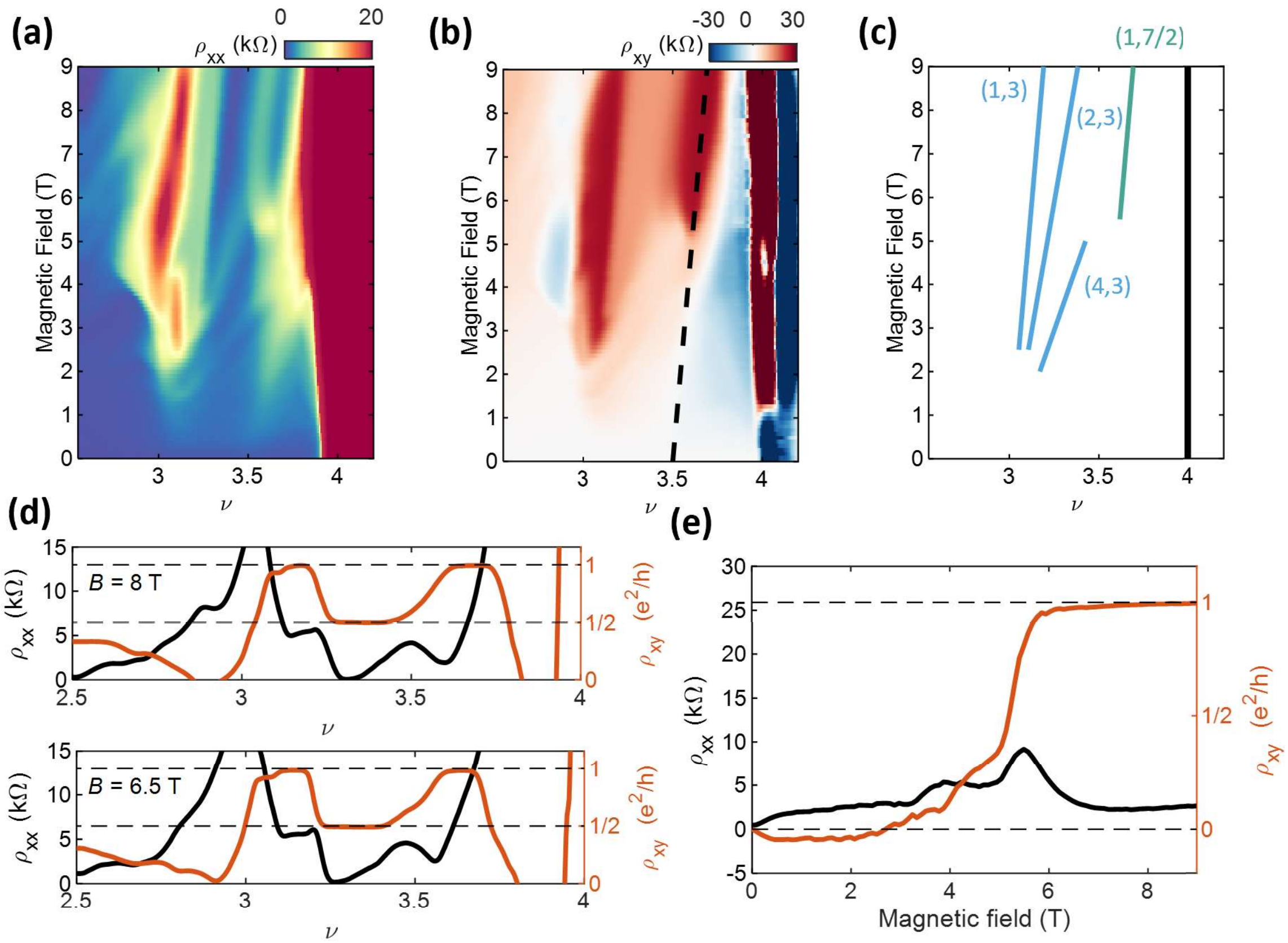

Figure S5: **Characterization of the (1, 7/2) SBCI. a-b**, Landau fan diagram of the longitudinal($\rho_{xx}$) and Hall($\rho_{xy}$) resistivity at $D$ = -0.4 V/nm, in the $\theta$' = 1.39° (AB-BA) tDBG device (Device O1). The data is restricted to focus on the correlated states surrounding $\nu = 3$ and 3.5. Data is acquired at 100 mK. **c**, Schematic of the most robust gapped states observed in **(a-b)**. The color scheme follows the convention detailed in Figs. 2-3 of the main text. **d**, Cuts of $\rho_{xx}$ and $\rho_{xy}$ at fixed magnetic field, $B$ = 8 T and $B$ = 6.5 T respectively. **e**, Cuts of $\rho_{xx}$ and $\rho_{xy}$ acquired along the trajectory of the (1, 7/2) SBCI, as shown by the black dashed line in **b**.

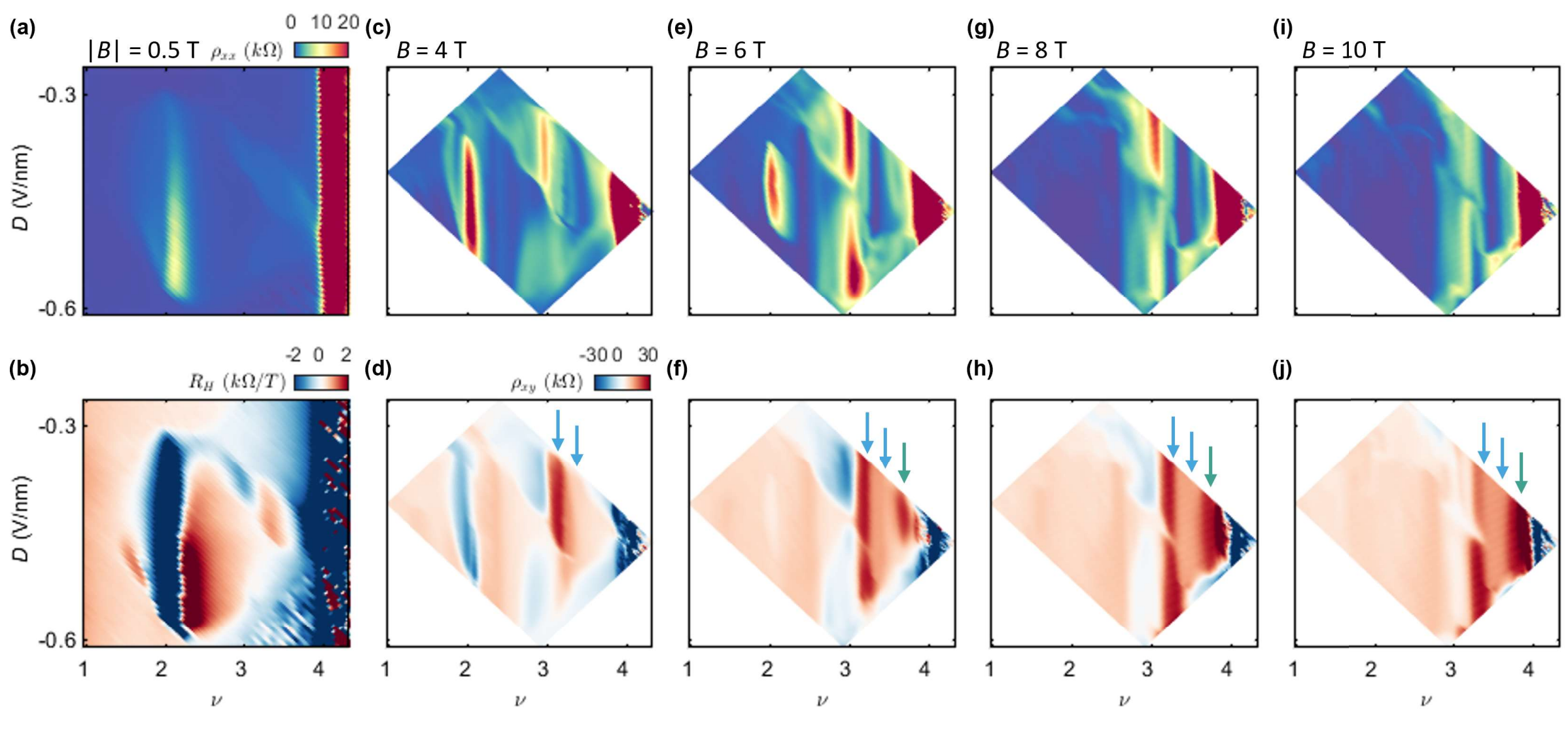

Figure S6: **Evolution of Chern insulators and the (1, 7/2) SBCI with magnetic field in the $\theta' = 1.39^\circ$ AB-BA tDBG device (O1). a-b**, Resistivity $\rho_{xx}$ and Hall coefficient $R_H$ map as a function of $D$ and $\nu$, acquired at $|B| = 0.5$ T. $\rho_{xx}$ and $R_H$ are field symmetrized and anti-symmetrized, respectively. **c-j**, $\rho_{xx}$ and $\rho_{xy}$ map at $B = 4$, 6, 8, and 10 T, respectively. Blue arrows denotes Chern insulators associated with s = 3. The green arrow denotes the (1, 7/2) SBCI, which emerges above 6 T. All data are acquired at 100 mK.

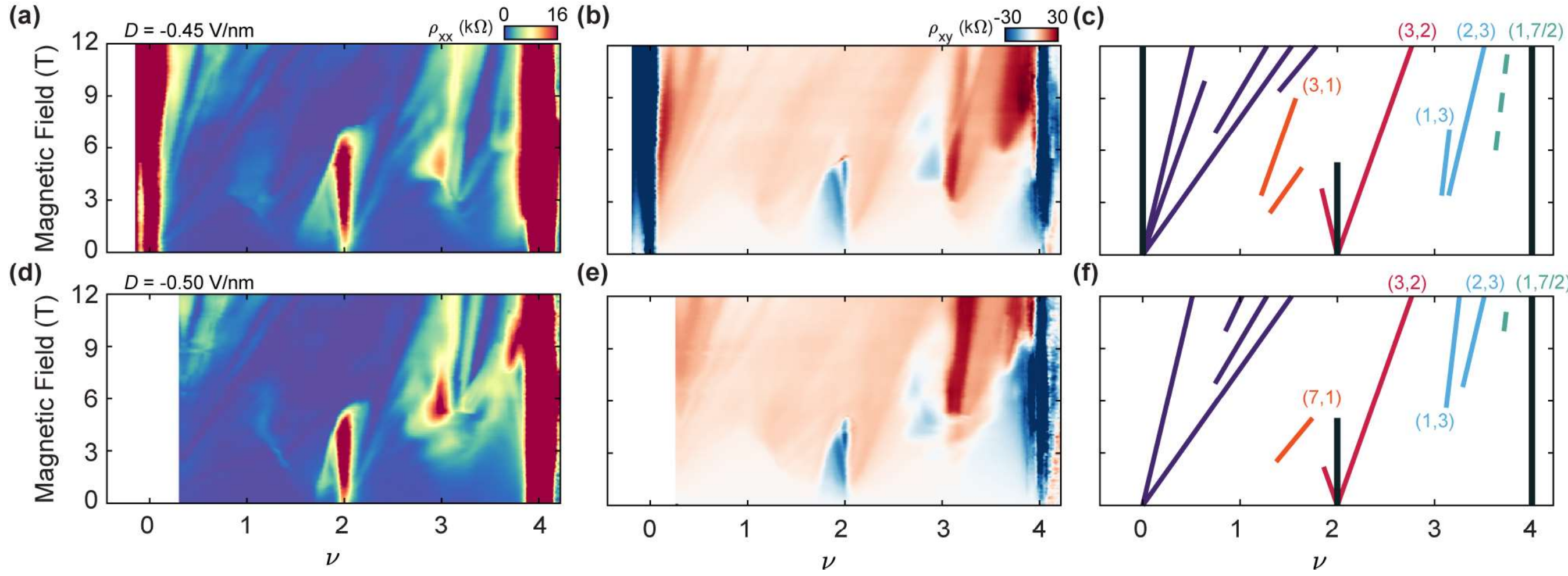

Figure S7: **Landau fan diagrams in the $\theta' = 1.39^\circ$ AB-BA tDBG device (O1) at selected $D$.** The left column shows the longitudinal resistivity ($\rho_{xx}$), the middle shows the Hall resistivity ($\rho_{xy}$), and right shows corresponding schematics of observed gapped states. $D$ = -0.45 V/nm for **a-c**, and $D$ = -0.50 V/nm for **d-f**. Note that **c** and **f** use the same color scheme as in Figs. 2-3 of the main text. All data are acquired at 100 mK.

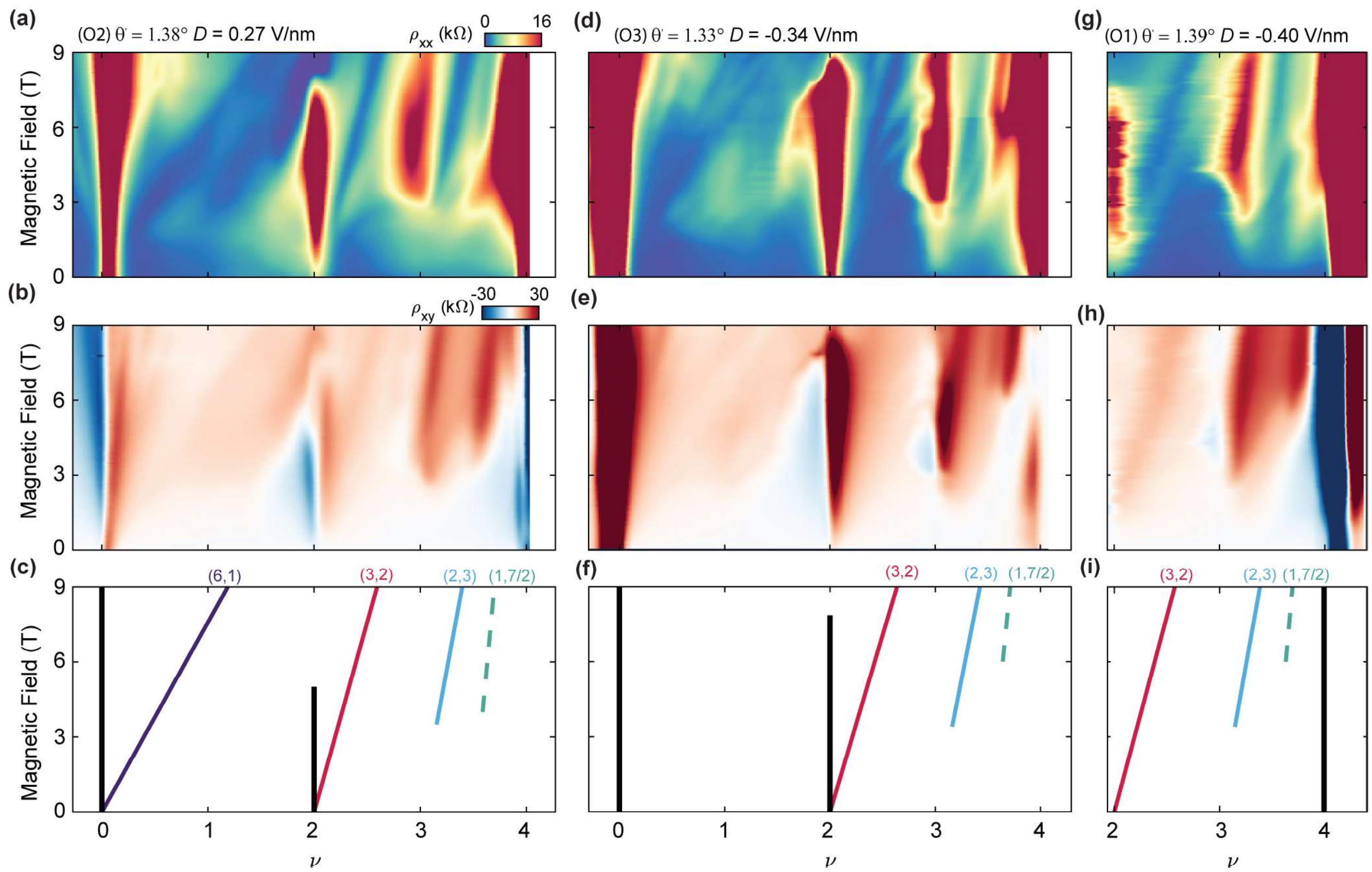

Figure S8: **Robust SBCI at $T = 2$ K observed in additional AB-BA tDBG devices.** Landau fan diagram in resistivity $\rho_{xx}$, Hall resistivity $\rho_{xy}$ and schematic of the observed gapped states for device (O2) $\theta' = 1.38°$ device in **a-c**, device (O3) $\theta' = 1.33°$ in **d-f**, and device (O1) $\theta' = 1.39°$ device in **g-i**. Note that **c,f,i** use the same color scheme as in Figs.2-3 of the main text. All data are acquired at $T = 2$ K. We find that the (1, 7/2) SBCI state as a robust ground state in AB-BA tDBG, and persists up to at least 2 K in all three devices shown.

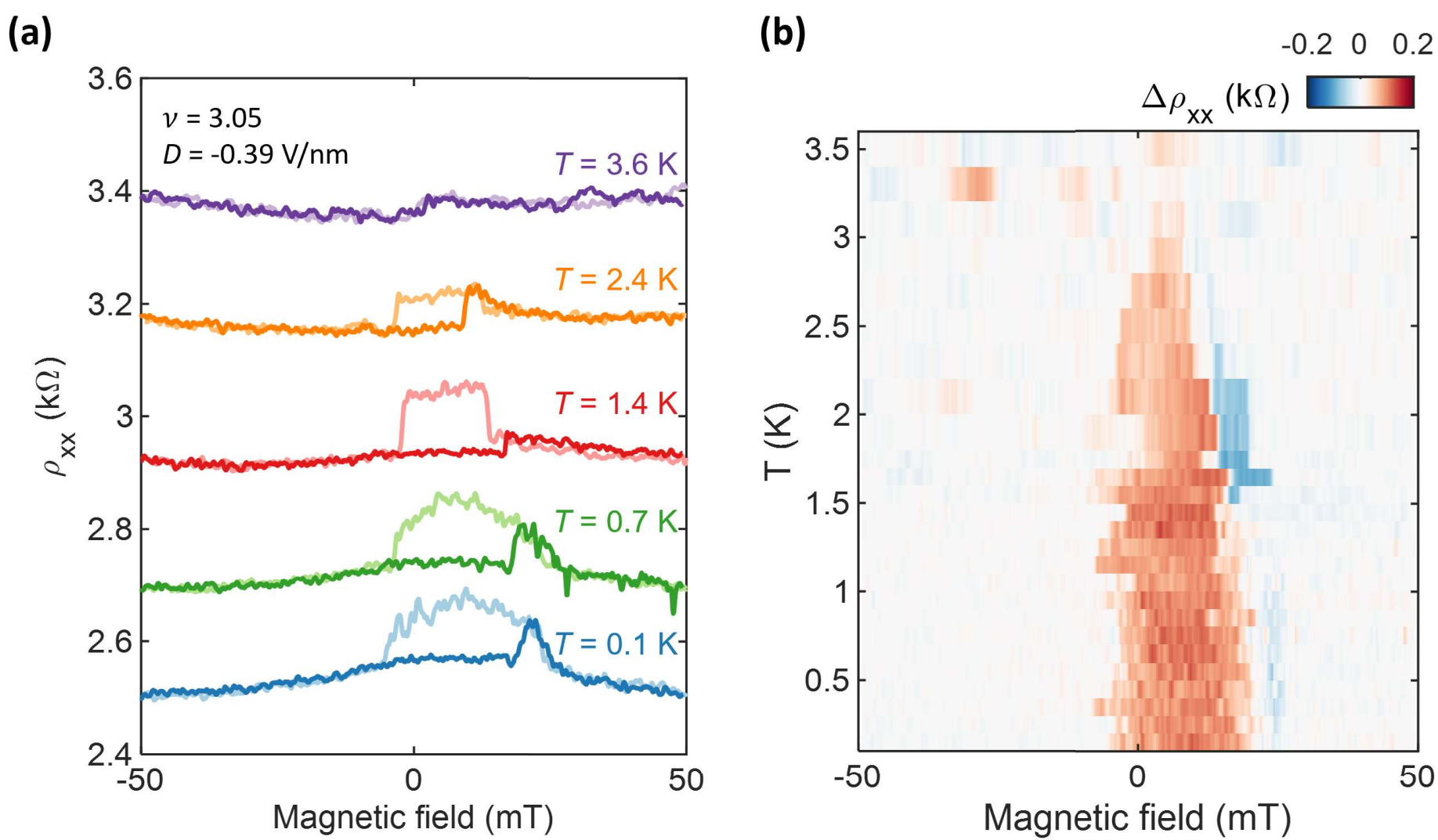

Figure S9: **Temperature dependence of the AHE near $\nu = 3$ in device O1.** **a**, $\rho_{xx}$ measured as magnetic field is swept back and forth at $\nu = 3.05$ and $D$ = -0.39 V/nm in the $\theta$' = 1.39° (AB-BA) tDBG device (Device O1). The curves are offset for clarity. Data are acquired in a separate cooldown from those presented in the main text, and subtle changes in the magnetic domain structure induced by thermal cycling result in a stronger AHE in the $\rho_{xx}$ geometry in this cooldown. This likely arises due to mixing between $\rho_{xx}$ and $\rho_{xy}$. **b**, $\Delta\rho_{xx} = \rho_{xx}^{B\downarrow} - \rho_{xx}^{B\uparrow}$ plotted as a function of magnetic field and temperature. The AHE vanishes above $\sim$ 3 K.

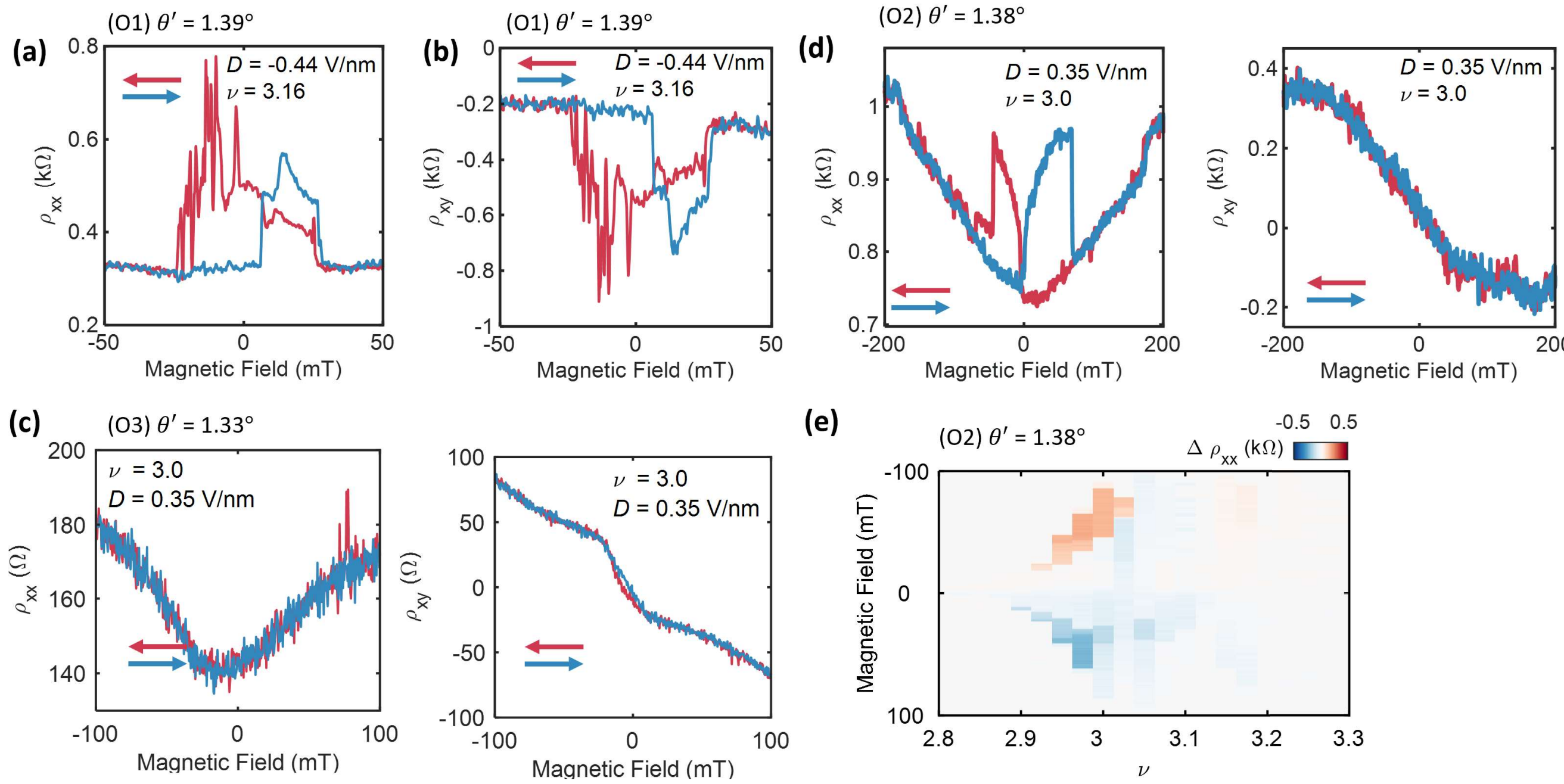

Figure S10: **AHE near $\nu = 3$ in three AB-BA tDBG devices. a-b**, $\rho_{xx}$ and $\rho_{xy}$ as the magnetic field is swept back and forth at specified $\nu$, $D$ in the $\theta' = 1.39°$ device (Device O1). Data is taken in a separate cooldown from those presented in Figs. 2d-e and Fig. S9. **c**, $\rho_{xx}$ and $\rho_{xy}$ as the magnetic field is swept back and forth at specified $\nu$, $D$ in the $\theta' = 1.33°$ device (Device O3). Data is acquired at 50 mK. $\rho_{xy}$ shows kink in the slope and a small hysteric loop at zero field, suggesting an emerging AHE at $\nu = 3$. **d**, $\rho_{xx}$ and $\rho_{xy}$ as field swept back and forth at specified $\nu$, $D$ in the $\theta' = 1.38°$ device (Device O2). **e**, $\Delta\rho_{xx}$ as a function of magnetic field measured at different filling factor with fixed $D = 0.35$ V/nm in the same device. Data are acquired at 300 mK in panels **d-e**.

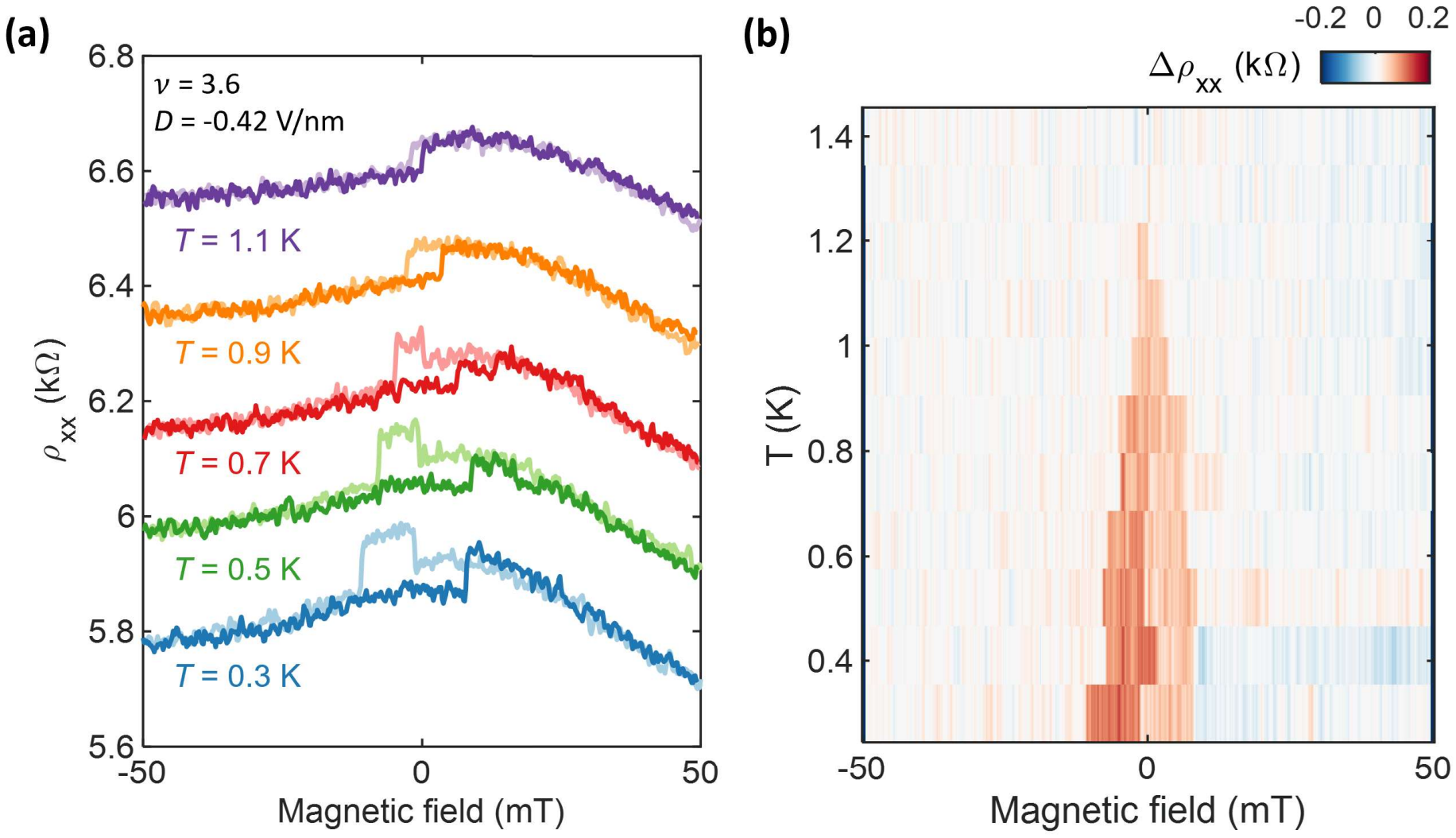

Figure S11: **Temperature dependence of the AHE near $\nu = 3.5$. a**, $\rho_{xx}$ measured as magnetic field is swept back and forth at $\nu = 3.6$ and $D$ = -0.42 V/nm in the $\theta$' = 1.39° (AB-BA) tDBG device (Device O1). The curves are offset for clarity. **b**, $\Delta\rho_{xx} = \rho_{xx}^{B\downarrow} - \rho_{xx}^{B\uparrow}$ plotted as a function of magnetic field and temperature. The AHE vanishes above $\sim$ 1.2 K.

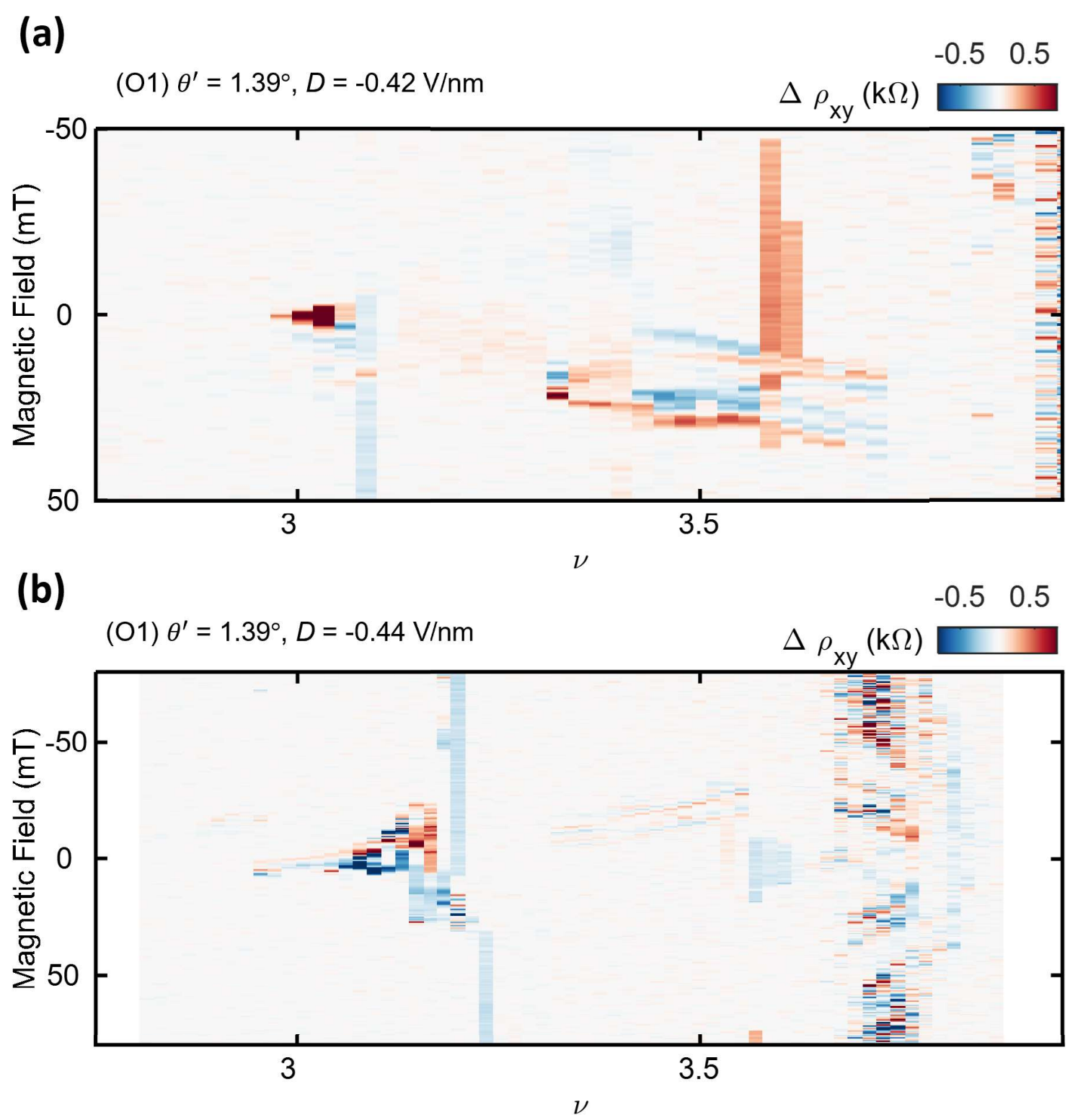

Figure S12: **Isolated AHE near $\nu = 3$ and $\nu = 7/2$.** $\Delta\rho_{xy}$ as a function of magnetic field measured at different filling factor at fixed **a** $D$ = -0.44 V/nm, **b** $D$ = -0.42 V/nm in the $\theta'$ = 1.39° AB-BA tDBG device (O1). We define $\Delta\rho_{xy} = \rho_{xy}^{B\downarrow} - \rho_{xy}^{B\uparrow}$ to visualize the AHE. The AHE at $\nu = 3$ and $\nu = 7/2$ are isolated, indicating they are likely independent correlated states. Data is acquired at 100 mK.

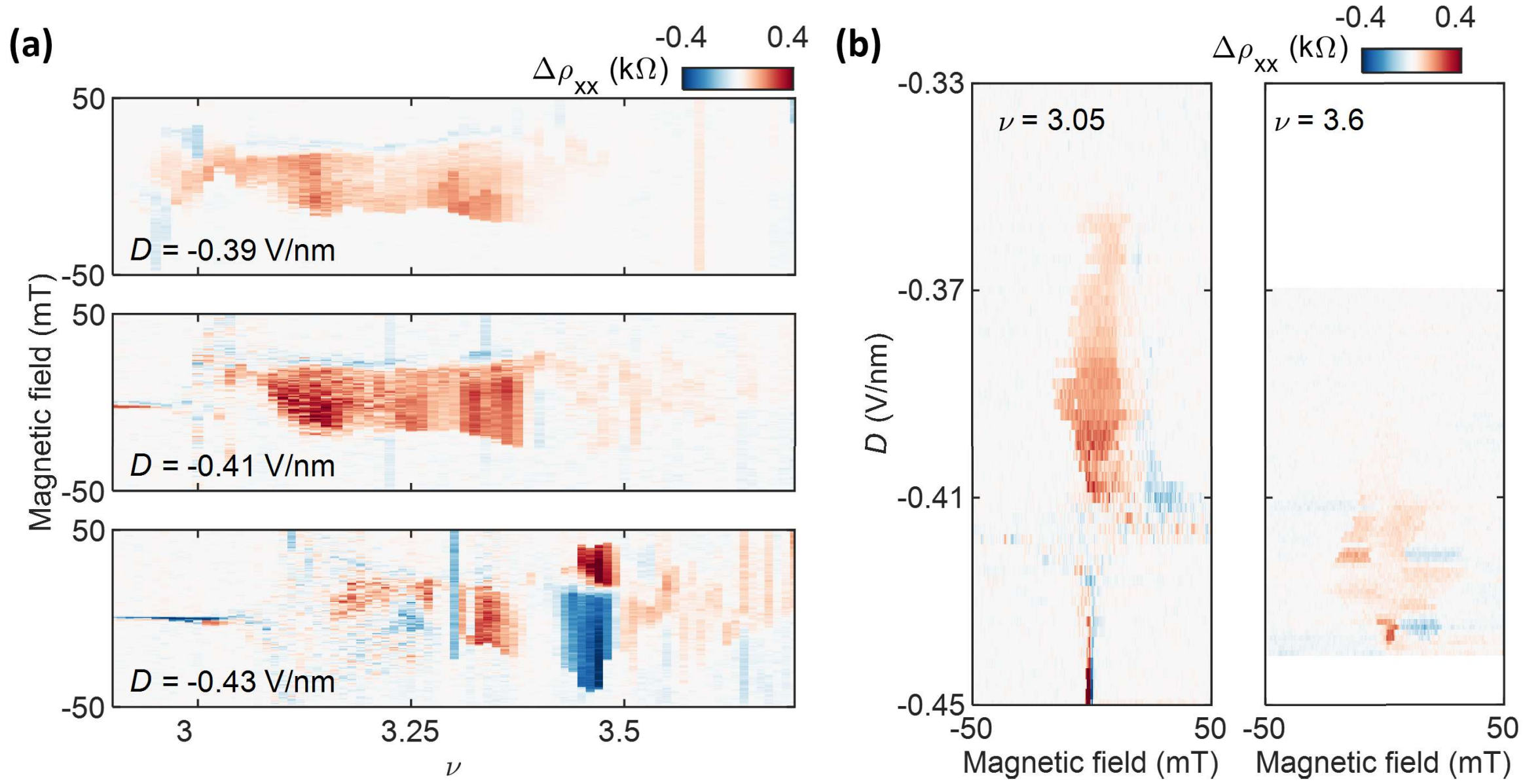

Figure S13: **Additional filling factor $\nu$ dependence and displacement field $D$ dependence of anomalous Hall effect.** **a**, $\Delta\rho_{xx} = \rho_{xx}^{B\downarrow} - \rho_{xx}^{B\uparrow}$ plotted as a function of magnetic field measured over a range of $\nu$ at fixed $D$ = -0.39, -0.41, -0.43 V/nm respectively, in the $\theta$' = 1.39° (AB-BA) tDBG device. Data are acquired in a separate cooldown from those presented in Figure S12, and subtle changes in the magnetic domain structure induced by thermal cycling result in a stronger AHE in the $\rho_{xx}$ geometry in this cooldown. The AHE near $\nu = 3$ and $\nu = 3.5$ appears less isolated but still separated from each other. **b**, $\Delta\rho_{xx} = \rho_{xx}^{B\downarrow} - \rho_{xx}^{B\uparrow}$ plotted as a function of magnetic field measured over a range of $D$ at fixed $\nu = 3.05$ and $\nu = 3.6$ respectively. All data is taken at 100 mK.

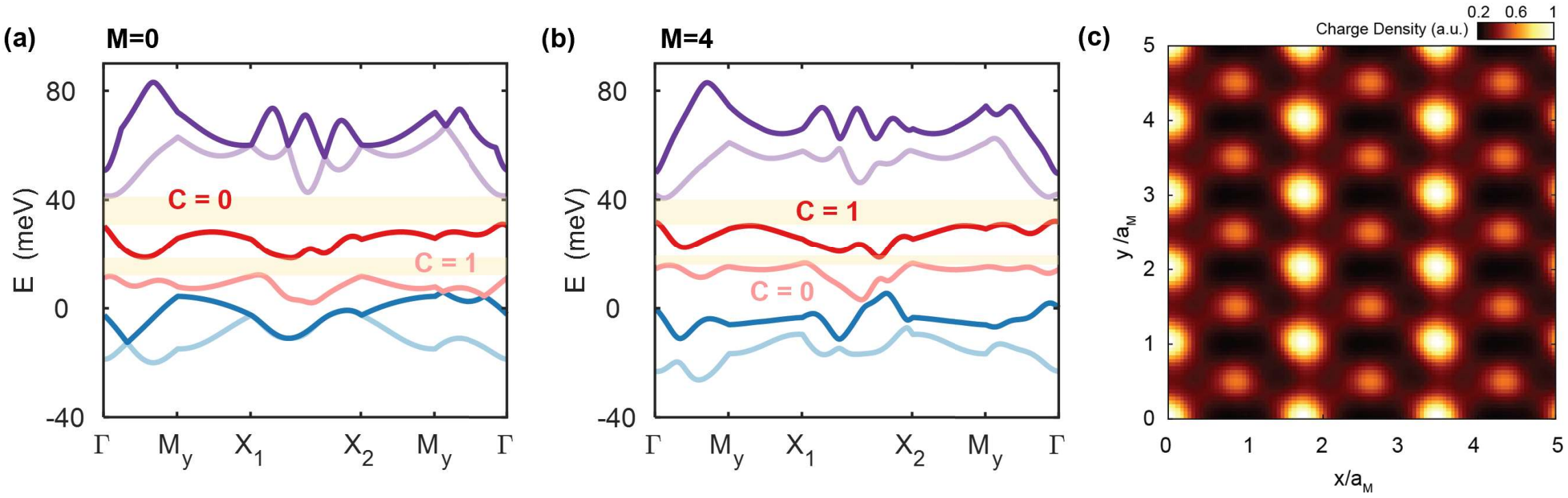

Figure S14: **Reconstructed band structure with CDW order.** Calculated band structure of AB-BA tDBG at $\theta' = 1.2°$, $\delta = 40$ meV, with charge density wave order $\Phi = 16$ meV. The red and blue bands denote the gapped subbands of the lowest moiré conduction and valence bands, respectively. The red subbands are split with a CDW gap of $\Delta = 2$ meV. The associated Chern numbers are labeled just above each conduction subband. We consider remote band renormalization with cutoff $M$, denoting number of remote bands considered. Results with M = 0 and M = 4 shown in **a,b** respectively. The partitioning of the Chern number of the conduction band is sensitive to $M$, and quickly converges when $M > 1$ to its real value (i.e., the value at $M = \infty$). **c** Calculated density profile in real space after introducing the CDW order. A gauge is chosen to pin the maximum of the CDW order to the AA sites. $a_M$ is the moiré lattice constant.

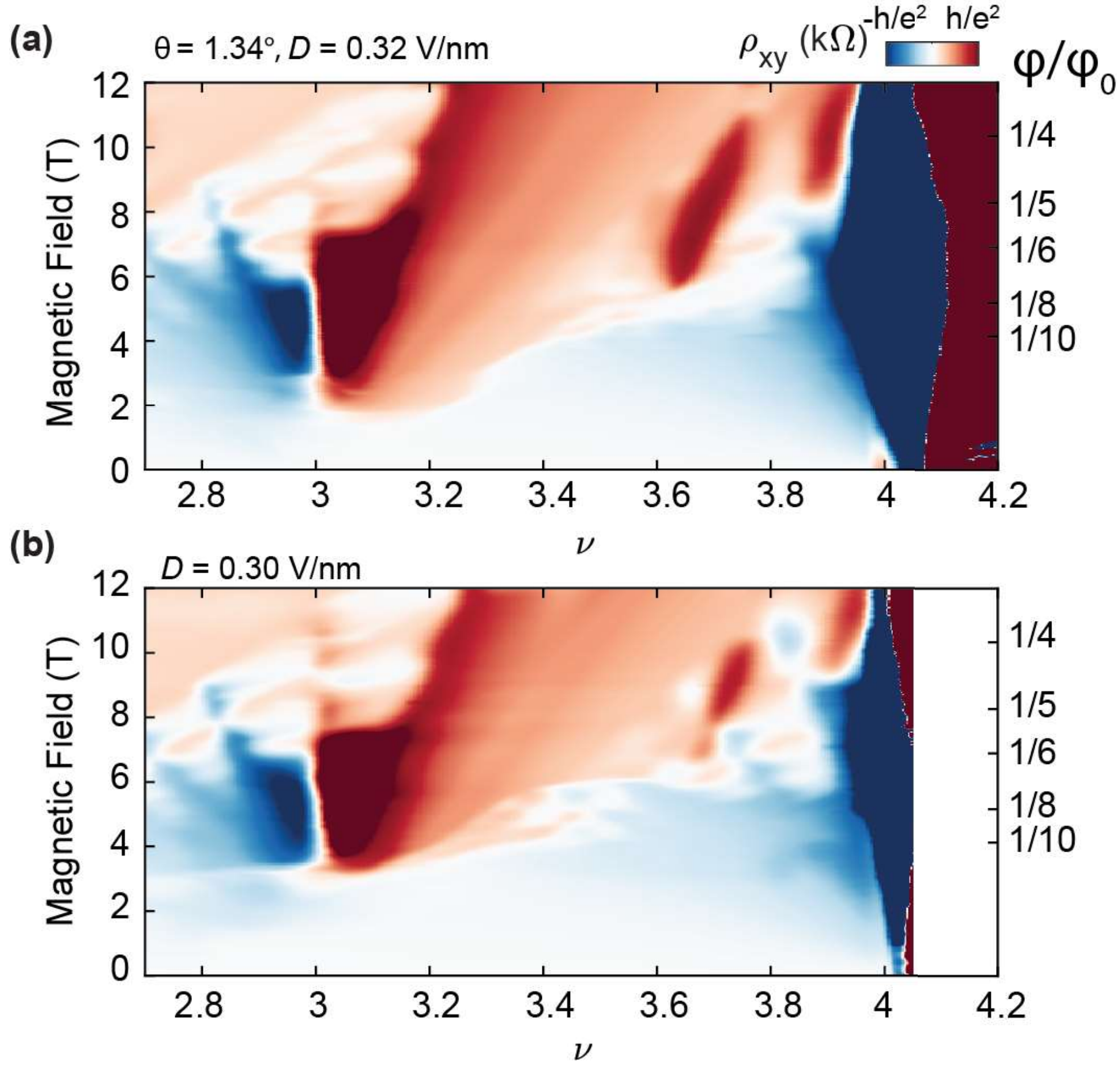

Figure S15: **SBCIs at other** $D$ **in device S2. a-b** Hall resistivity $\rho_{xy}$ of the $\theta = 1.34°$ AB-AB tDBG (device S2), at different $D$. Both Landau fan features the (1, 7/2) and the (1, 11/3) symmetry broken Chern insulators yet with subtle $D$ dependence of its optimal $B$ field range.

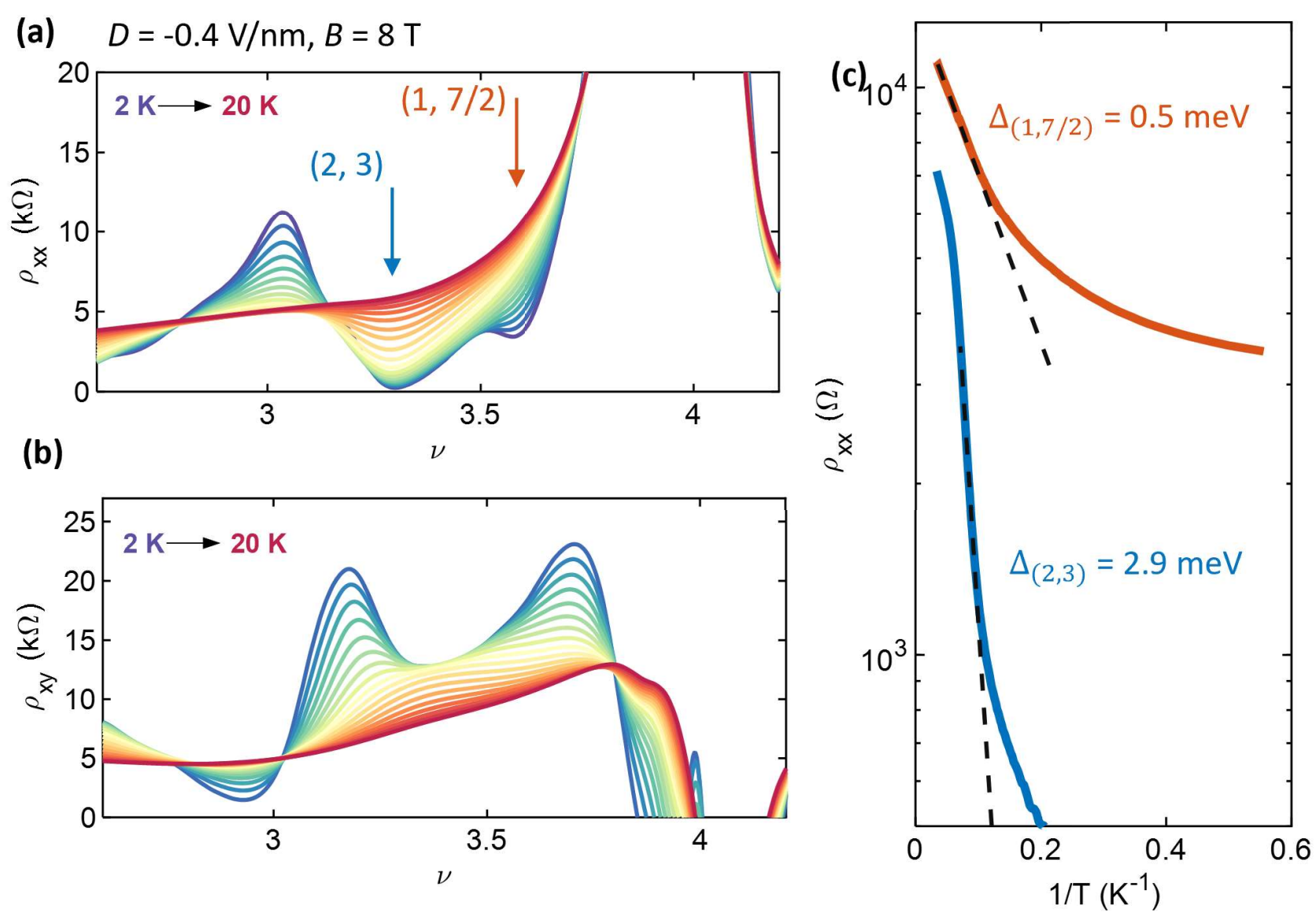

Figure S16: **Temperature dependence and energy gaps of the Chern insulators in device O1. a-b**, Longitudinal resistivity $\rho_{xx}$ and Hall resistivity $\rho_{xy}$ of the $\theta' = 1.39^\circ$ AB-BA tDBG (device O1), respectively, plotted as a function of $\nu$ measured at different temperatures in steps of 1 K. The suppression in $\rho_{xx}$ is characteristic feature of the (2, 3) correlated Chern insulator and the (1, 7/2) symmetry broken Chern insulator. **c**, Arrhenius plots of the $\rho_{xx}$ at the (2, 3) and (1, 7/2) states (blue and orange, respectively). In the quantum Hall effect (small $\sigma_{xx}$ limit), we have $\rho_{xx} = \sigma_{xx}/(\sigma_{xx}^2 + \sigma_{xy}^2) \propto \sigma_{xx} \propto e^{-\Delta/2T}$. The dashed lines denote representative activation fits, which correspond to an energy gap $\Delta =$ 0.5 meV for the (1, 7/2) SBCI and $\Delta = 2.9$ meV for the (2, 3) CI.

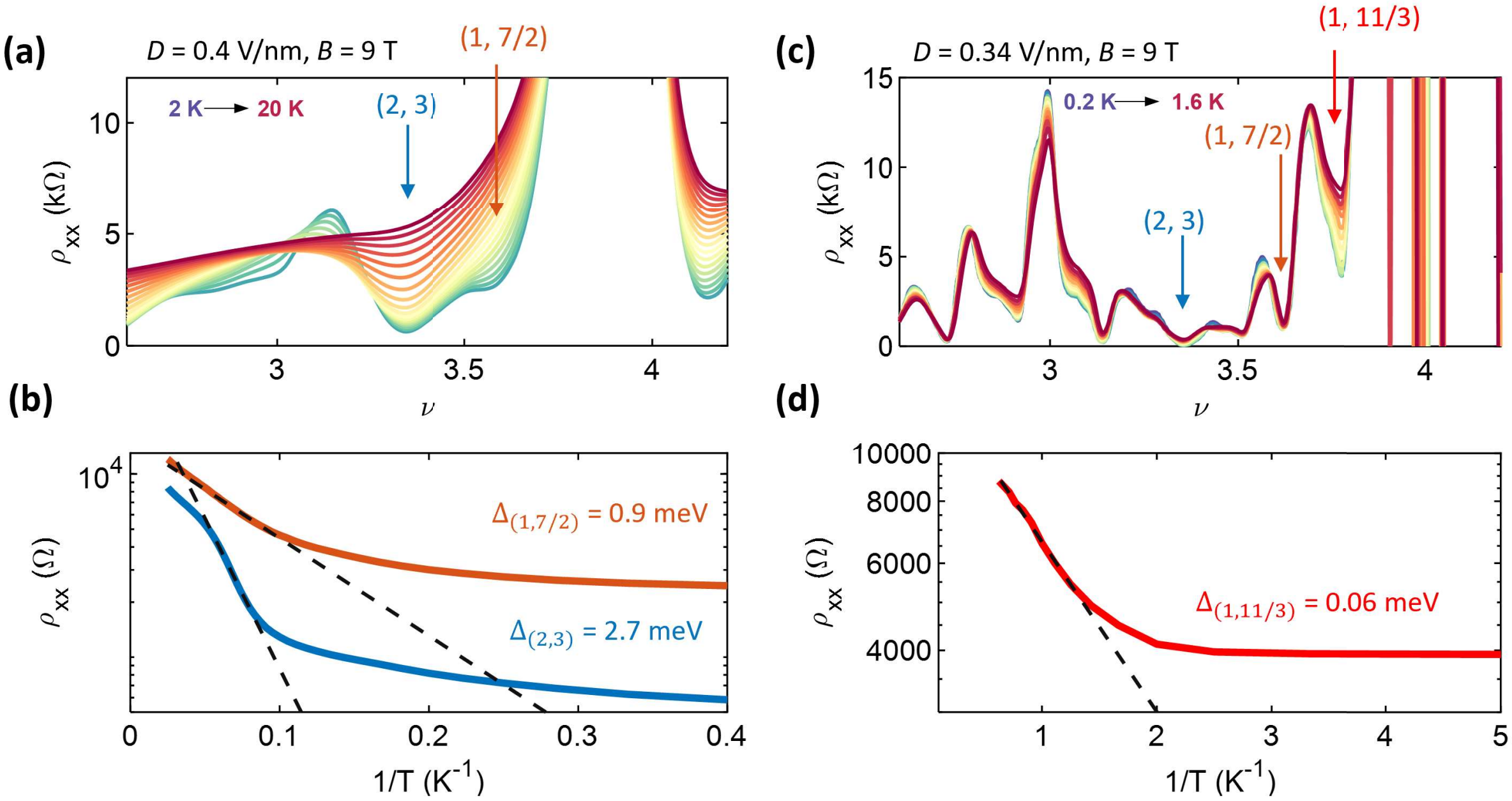

Figure S17: **Temperature dependence and energy gaps of the Chern insulators in device S2.** **a**, Longitudinal resistivity $\rho_{xx}$ of the $\theta = 1.34^{\circ}$ AB-AB tDBG (device S2), plotted as a function of $\nu$ measured at different temperatures in steps of 1 K from 2K to 20K. The suppression in $\rho_{xx}$ is characteristic feature of the (2, 3) correlated Chern insulator and the (1, 7/2) symmetry broken Chern insulator. **b**, Arrhenius plots of the $\rho_{xx}$ at the (2, 3) and (1, 7/2) states (blue and orange, respectively). We fit an energy gap $\Delta = 0.9$ meV for the (1, 7/2) SBCI and $\Delta = 2.7$ meV for the (2, 3) CI. **c**, Longitudinal resistivity $\rho_{xx}$ of the same device S2 measured at lower temperature in steps of 100 mK from 200 mK to 1.6K. The suppression in $\rho_{xx}$ is characteristic feature of the (1, 11/3) SBCI. **d**, Arrhenius plots of the $\rho_{xx}$ at the (1, 11/3). We fit an energy gap $\Delta = 0.06$ meV for the (1, 11/3) SBCI.

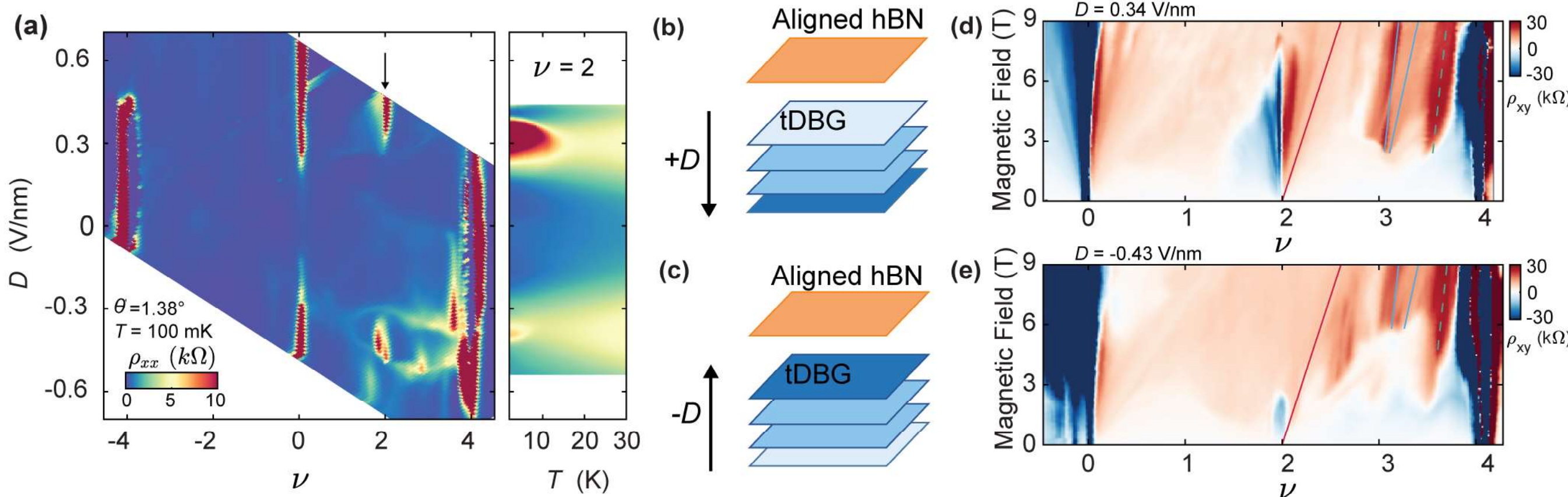

Figure S18: **Robust SBCI with additional top hBN alignment. a,b**, Resistivity ($\rho_{xx}$) map of a $\theta' = 1.38^\circ$ AB-BA tDBG device aligned with top hBN at a twist angle of $0.55^\circ$. The right panel shows $\rho_{xx}$ as function of temperature and $D$ at $\nu = 2$. **b-c**, Schematics of the layer-resolved electron density upon a positive and negative $D$, respectively. The electron density is schematically indicated by intensity of blue coloring. **d-e**, Hall resistivity, $\rho_{xy}$, measured at $D = 0.34$ V/nm (top) and $D =$ -0.43 V/nm (bottom), respectively. Schematics of observed gapped states are overlayed, using the same color coding as in maintext Fig. 3.

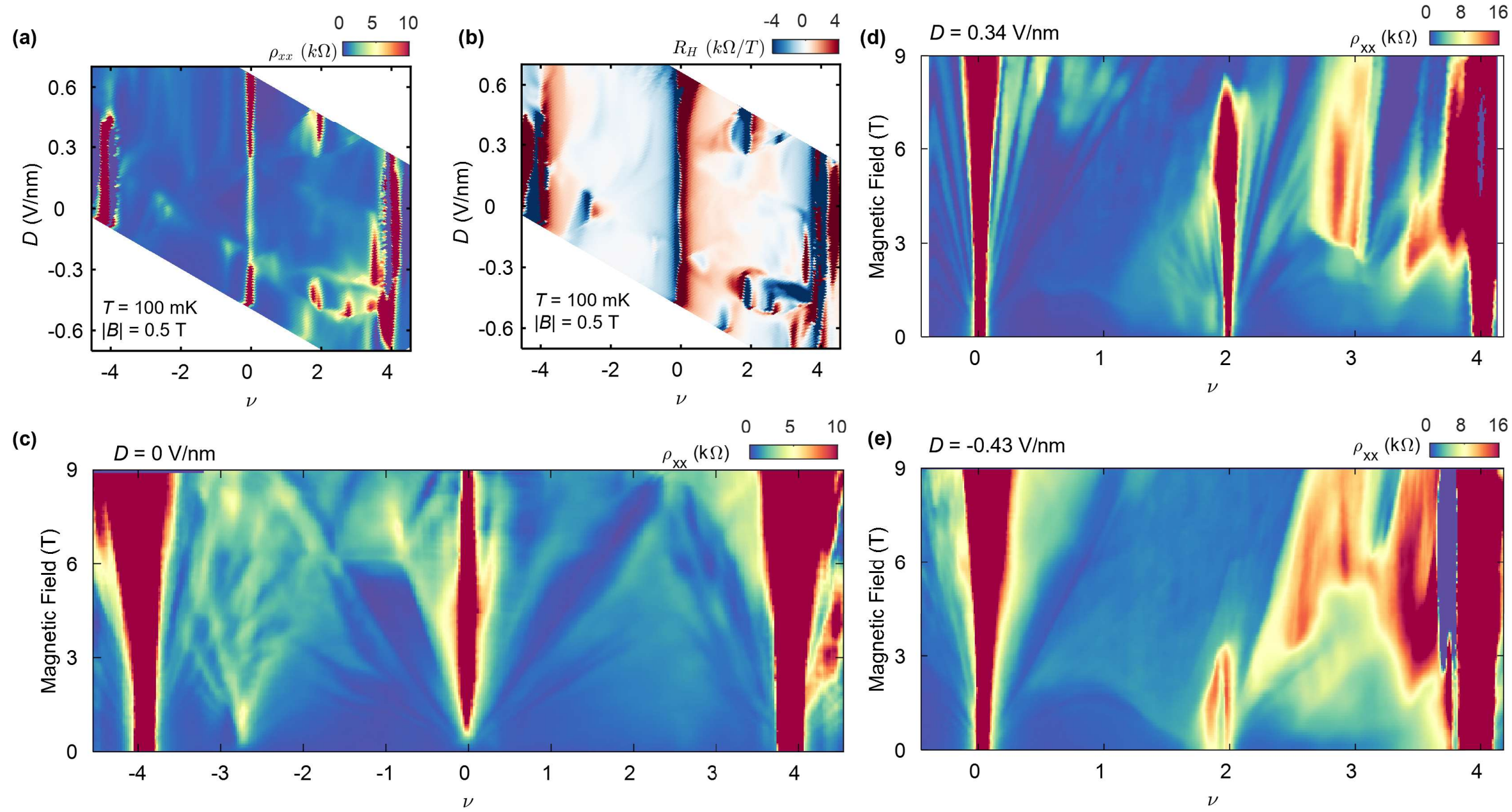

Figure S19: $\theta' = \mathbf{1.38^\circ}$ **AB-BA tDBG device aligned with hBN. a,b**, Resistivity $\rho_{xx}$ and Hall coefficient $R_H$ map, field symmetrized and anti-symmetrized with $|B| = 0.5$ T, respectively. **c**, Landau fan diagram of the longitudinal resistivity $\rho_{xx}$ at $D = 0$ V/nm. Additional set of gapped states are seen emerging from $\nu \approx -2.8$, corresponding to the secondary Dirac point from the graphene/hBN moiré. We calculate the twist angle between graphene and hBN to be 0.55° based on the carrier density of the secondary Dirac point. **d-e**, Landau fan of $\rho_{xx}$ at $D = +0.34$ V/nm and $D = -0.43$ V/nm, respectively. The gapped states observed in the Landau fans vary substantially with the sign of $D$ owing to the alignment of the hBN on one side. All data are acquired at 100 mK.

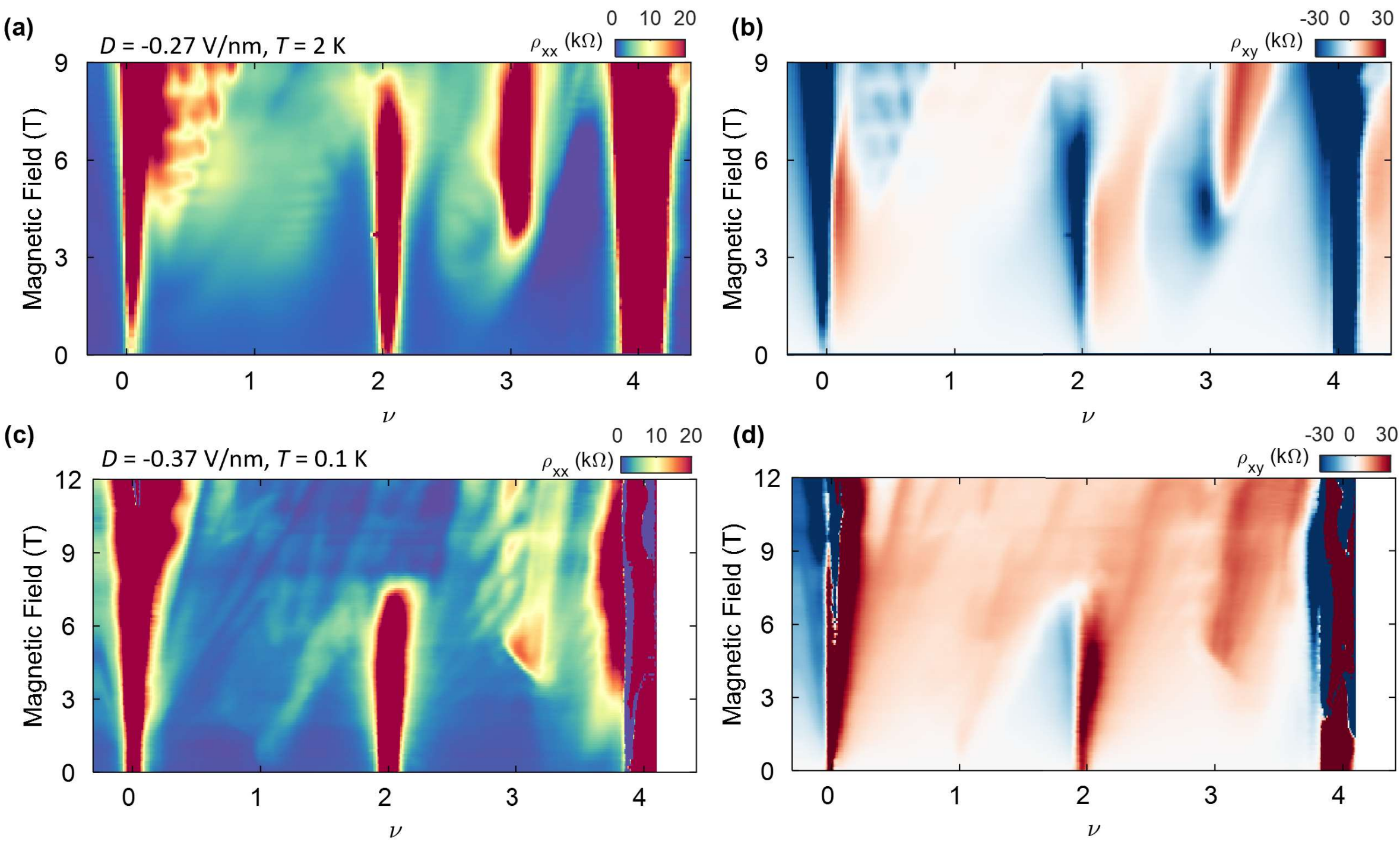

Figure S20: **Absence of the (1, 7/2) SBCI in $\theta$ = 1.30° tDBG.** Landau fan of the longitudinal resistivity, $\rho_{xx}$, and Hall resistivity, $\rho_{xy}$, measured at **a-b**, $D$ = -0.27 V/nm, $T$ = 2 K. **c-d**, $D$ = -0.37 V/nm, $T$ = 0.1 K in a $\theta$ = 1.30° AB-AB tDBG (device S2). We do not observe a SBCI state at this twist angle.

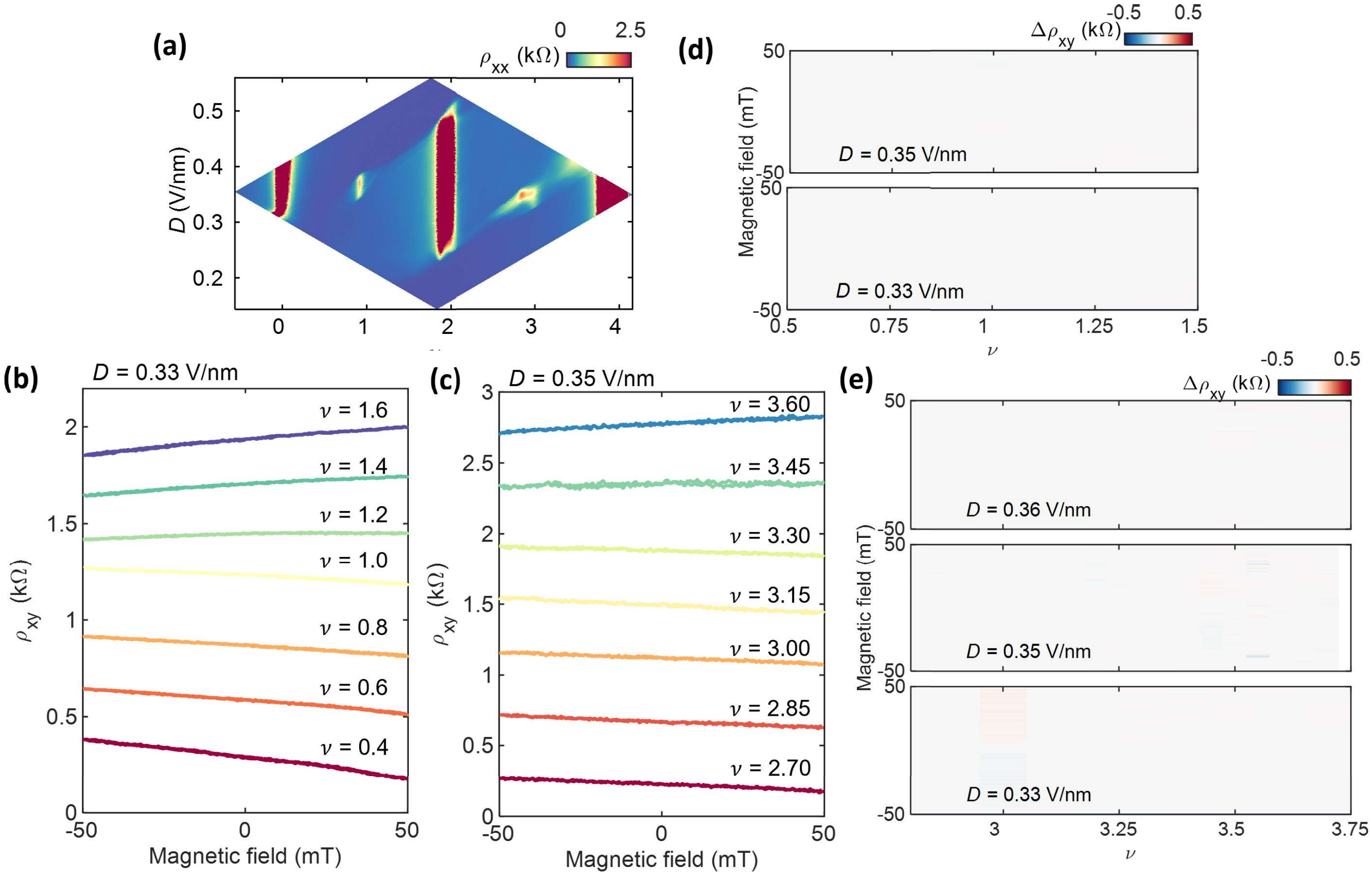

Figure S21: **Absence of the AHE in $\theta = 1.30°$ AB-AB tDBG. a**, $\rho_{xx}$ map of a $\theta = 1.30°$ AB-AB stacked tDBG sample acquired at 100 mK. We observe weak correlated insulating states at $\nu = 1$ and 3. **b-c**, Representative measurements of $\rho_{xy}$ in which the magnetic field is swept back and forth at fixed $D$, acquired at various values of $\nu$, surrounding $\nu = 1$ (panel b) and 3 (panel c). The curves are offset for clarity. We observe a linear (ordinary) Hall effect at all $\nu$, and well as a characteristic reversal in the sign of the slope of the Hall effect upon doping across integer filling. **d-e**, $\Delta\rho_{xy} = \rho_{xy}^{B\downarrow} - \rho_{xy}^{B\uparrow}$ plotted as a function of magnetic field measured over a small range of $\nu$ surrounding 1 and 3, acquired at fixed $D$ as indicated in the plots. Data are all acquired at 100 mK. We observe no signatures of hysteresis associated with the AHE at this particular twist angle.